%% file: 00_Main.tex
\documentclass[lettersize,journal]{IEEEtran}
\usepackage{amsmath,amsfonts}
\usepackage{dsfont}
\usepackage{algorithmic}
\usepackage{algorithm}
\usepackage{array}
\usepackage[caption=false,font=normalsize,labelfont=sf,textfont=sf]{subfig}
\usepackage{textcomp}
\usepackage{stfloats}
\usepackage{url}
\usepackage{verbatim}
\usepackage{graphicx}
\usepackage{epstopdf}
\usepackage{tabularx}
\usepackage{cite}
\usepackage{flushend}
\usepackage{tikz}
\usepackage{dirtytalk}
\newcommand{\Gaussian}{\mathcal{N}}
\newcommand{\aux}{\mathcal{L}}
\newcommand{\vect}[1]{{#1}}
\newcommand{\mathintext}[1]{$#1$}
\newcommand{\timeind}{k}

\newcommand{\targetind}{m}
\newcommand{\Targetind}{M_\timeind}
\newcommand{\pixelindx}{i}
\newcommand{\Pixelindx}{I}

\newcommand{\pixelind}{i}
\newcommand{\Pixelind}{I}
\newcommand{\itind}{p}

\newcommand{\transp}{\mathrm{\textsf{T}}}
\newcommand{\makesubandsuper}[3]{#1_{#2}^{#3}}

\begin{document}

\title{A Scalable Hybrid Track-Before-Detect Tracking System: Application to Coastal Maritime Radar Surveillance}

\author{Lukas~Herrmann,
        Ángel~F.~García-Fernández,
        Edmund~F.~Brekke,
        and Egil Eide
\thanks{This work was supported by the Norwegian Research Council under SFI AutoShip (project number 309230), and the European Union’s Horizon 2020 research and innovation program under the PERSEUS doctoral program (Marie Skłodowska-Curie grant agreement number 101034240).}
\thanks{Lukas Herrmann and Egil Eide are with the Department of Electronic Systems, Norwegian
University of Science and Technology, Trondheim, Norway (email: lukas.herrmann@ntnu.no; egil.eide@ntnu.no).}
\thanks{Ángel F. Garc\'ia-Fern\'andez is with ETSI de Telecomunicaci\'on, Universidad Polit\'ecnica de Madrid, 28040 Madrid, Spain (email: \mbox{angel.garcia.fernandez@upm.es}).}
\thanks{Edmund F. Brekke is with the Department of Engineering Cybernetics, Norwegian
University of Science and Technology, Trondheim, Norway (email: edmund.brekke@ntnu.no).}}



\maketitle

\begin{abstract}
Despite their theoretical advantages, track-before-detect (TBD) methods remain largely absent from real-world multi-target tracking applications due to their computational complexity and limited scalability. This paper presents a scalable hybrid tracking framework that combines a TBD multi-target tracking algorithm with a detection-based multi-target tracking algorithm for coastal radar surveillance. In particular, the approach uses an integrated existence Poisson histogram-probabilistic multi-hypothesis tracking (IE-PHPMHT)-based TBD module with a conventional Poisson multi-Bernoulli Mixture (PMBM) point tracker. The system processes raw radar data through land clutter suppression, cell-wise detection, and clustering-based feature extraction. High-threshold detections are used to track strong targets via the point tracker, while low-threshold detections are employed for adaptive birth in the TBD module, enabling early initiation and sustained tracking of weak or ambiguous targets. Validated using real X-band radar data from the Trondheim Fjord, Norway, the approach demonstrates robust multi-target tracking performance in a full-scale application with a large observation area under resource constraints, highlighting its suitability for operational deployment in complex maritime environments needed for coastal surveillance and to support autonomy.
\end{abstract}


\input{01_Introduction}
\input{02_Problem_System}
\input{03_Framework}
\input{04_Imp_ExpEval}

\input{05_Conclusion}

\section*{Acknowledgment}
We would like to thank Nicholas Dalhaug, Ernesto López, and Henrik Flemmen for their assistance in operating the two target vessels during the experiments, and SINTEF Ocean AS and Eelume AS for making the vessels available.


\input{00_Main.bbl}

\end{document}

%% file: 01_Introduction.tex
\section{Introduction}
\label{sec:Introduction}

\IEEEPARstart{S}{hore}-based radar surveillance systems are essential components of modern maritime \textit{situational awareness} (SITAW) architectures, providing continuous, all-weather observation capabilities for coastal waters \cite{maresca_maritime_2014,vivone_joint_2016}. These systems contribute significantly to maritime safety, security, and operational efficiency by enabling the persistent detection, tracking, and classification of vessels in real time. With the increasing complexity of maritime operations, including dense traffic near critical infrastructure, and the emergence of autonomous vessels, the reliance on coastal monitoring systems is crucial \cite{forti_next-gen_2022}.

In the context of maritime operations, situational awareness supports early identification of vessel behaviours, recognition of potential navigational hazards, and coordination among assets. This capability becomes especially critical when facilitating autonomous maritime operations, wherein decision-making must be informed by timely and accurate environmental perception~\cite{brekke_autosea_2019}.
Autonomous maritime platforms, such as \textit{unmanned surface vessels} (USVs) and remotely operated systems, require a reliable external information source to augment onboard sensing limitations. Shore-based radar systems serve as a reliable source of SITAW. These systems can deliver kinematic and contextual data, such as vessel trajectories, into autonomous navigation stacks.
In addition to enabling autonomy, shore-based radar systems support a diverse range of mission-critical applications such as search and rescue, securing waterways and critical infrastructure protection, environmental protection, and navigation safety.





Conventional radar tracking pipelines typically rely on the detect-then-track paradigm \cite{bar-shalom_estimation_2001, richards_fundamentals_2014}, in which thresholded detections are passed to a tracking algorithm such as multi-hypothesis tracking \cite{reid_algorithm_1979, mellema_improved_2020}, probabilistic data association \cite{fortmann_multi-target_1980, fortmann_sonar_1983}, or \textit{random finite set} (RFS)-based methods \cite{mahler_advances_2014} such as the \textit{Poisson multi-Bernoulli mixture} (PMBM) filter \cite{williams_marginal_2015,garcia-fernandez_poisson_2018}. While this paradigm is widely applied and can show excellent tracking performance, a drawback is that there is an inherent loss of information in the thresholding operation.

The \textit{track-before-detect} (TBD) paradigm addresses this drawback by operating directly on unthresholded data to make use of all possible information available~\cite{mallick_integrated_2013}. This improves sensitivity, enabling earlier track initiation and sustained tracking of targets with low \textit{radar cross sections} (RCSs) or when embedded in strong clutter \cite{davey_comparison_2008}. 
TBD algorithms have been proposed based on RFSs \cite{davies_information_2024, garcia-fernandez_track-before-detect_2016, ristic_bernoulli_2020, vo_joint_2010}, dynamic programming\cite{barniv_dynamic_1985, yi_efficient_2013}, particle filters \cite{salmond_particle_2001, rutten_recursive_2005, morelande_bayesian_2007, northardt_track-before-detect_2019}, or \textit{histogram-probabilistic multi-hypothesis tracking} (H-PMHT) \cite{streit_tracking_2000, davey_track-before-detect_2018}.
H-PMHT has gained recognition for its comparably low computational complexity and, therefore, its potential real-time capability~\cite{davey_comparison_2008}. In a recent development, the \textit{integrated existence} Poisson H-PMHT (IE-PHPMHT), which incorporates an existence-based framework through a Bernoulli target representation, has been proposed to handle an unknown and time-varying number of targets~\cite{herrmann_histogram-probabilistic_2025}.

In prior work, a few studies have applied TBD techniques to real sensor data. Airborne radar data have been used for single-target tracking analysis of a small boat in \cite{mcdonald_track-before-detect_2008,mcdonald_impact_2011}, and the same dataset was subsequently employed to demonstrate the H-PMHT algorithm in \cite{davey_detecting_2011}. An extension of H-PMHT accounting for sensor location uncertainty was presented in \cite{guo_sa-hpmht_2023}. In a different line of work, correlation filtering techniques were applied to maritime vessel tracking in \cite{zhou_multiple-kernelized-correlation-filter-based_2022}, and most recently, a qualitative proof of concept of a multi-Bernoulli TBD approach for airborne maritime radar was shown in \cite{ristic_track-before-detect_2024}.

Despite considerable progress in algorithm development, the vast majority of TBD methods have been evaluated primarily under simulated conditions.
In addition, in most cases, these studies are constrained to a small amount of resolution cells.
While simulation results have demonstrated the potential of TBD to markedly enhance detection and tracking performance, the transition to real-world applications has been limited. This gap is primarily due to the significant computational burden associated with TBD, especially in the context of multiple targets across wide surveillance areas.





In this paper, we propose a scalable solution to multi-target TBD tracking for coastal maritime radar surveillance that achieves both improved tracking performance and real-time capability. This is accomplished by utilising IE-PHPMHT and is further supported by the development of a hybrid tracking architecture composed of a PMBM point tracker combined with the IE-PHPMHT track-before-detect approach. The entire signal processing chain includes preprocessing in the form of land clutter removal, applying a cell-wise detector, and feature extraction employing clustering. The obtained point detections with a high threshold are provided to the PMBM point tracker to track high \textit{signal-to-noise ratio} (SNR) targets such as ferries, cargos, and cruise ships. Conversely, for the low SNR targets, the TBD module runs in parallel and low threshold detections are used for the adaptive birth \cite{herrmann_track_2025} within IE-PHPMHT for potential track initiation. Finally, the output of the hybrid tracking system is the combined output of both trackers.

The proposed solution is validated via its application to a real data set acquired by an X-band marine radar observing a substantial part of the Trondheim Fjord, Norway, including the Trondheim harbour area.
The proposed solution demonstrates, first, the successful implementation of a TBD method, the IE-PHPMHT, in a real-world application, i.e. multiple appearing and disappearing targets in a large observation area (processing radar images with up to 2048 $\times$ 720 resolution cells) with limited computational resources. Second, the development of a hybrid tracking framework that fuses a conventional detection-based point-tracking subsystem with a TBD module, which allows the system to scale efficiently while preserving the benefits of TBD for challenging detections. Third, the full-scale demonstration and evaluation of the tracking system through the deployment of a field test, processing continuous radar data and performance assessment using \textit{global navigation satellite system} (GNSS) and \textit{automatic identification system} (AIS) ground truth data.

The paper is organised as follows. Section \ref{sec:Prob} gives a system overview and contextualises the problem formulation. The entire tracking framework, including the single steps of the signal processing chain, is described in Section \ref{sec:framework} while Section \ref{sec:Imp_Eval} provides the real data results and the performance assessment. Finally, Section \ref{sec:conclusion} concludes the paper.

%% file: 02_Problem_System.tex
\section{Problem Statement and System Overview}
\label{sec:Prob}

We address the problem of multi-target tracking in a littoral environment using a radar system. The surveillance scenario involves monitoring vessel activity in the Trondheim Fjord, a large and coastal area characterised by mixed maritime traffic, variable sea states, and static land clutter. 

Let the multi-target state at discrete time $k \in \{1,...,K\}$ be represented $X_k = \left[ x_k^{1},\dots, x_k^{m}, \dots, x_k^{M_k} \right]^\transp$, where $m$ ranges over the index set $\mathcal{M}_\timeind = \left\{1, \ldots, \Targetind\right\}$ and $M_k \in \mathbb{N}$ represents the unknown and time-varying number of targets. Each individual target state is given by $x_\timeind^{m}~=~\left[ p_{x,\timeind}^{m} \ v_{x,\timeind}^{m} \ p_{y,\timeind}^{m} \ v_{y,\timeind}^{m} \right]^{\!\transp}~\in~\mathbb{R}^4 $ comprising the position and velocity components in both the $x$- and $y$-directions.
At each time step $k$, the radar system produces measurements $Z_k = \left[ z_k^1,..., z_k^i,...,z_k^I \right]^\transp$ where each $z_k^i$ denotes the received power in the $i$-th range-azimuth cell, and $I$ is the total number of resolution cells.
The explicit dynamic and measurement models for the two different tracking algorithms are specified in their respective Sections \ref{sec:IE_PHPMHT} and \ref{sec:PMBM}.
The objective is to estimate the evolving multi-target state $X_k$ over time, based on the accumulated measurement history $Z_{1:k} = \left( Z_1,\dots,Z_k \right)$.

\begin{figure}[t]
    \centering
    \includegraphics[width=0.95\linewidth]{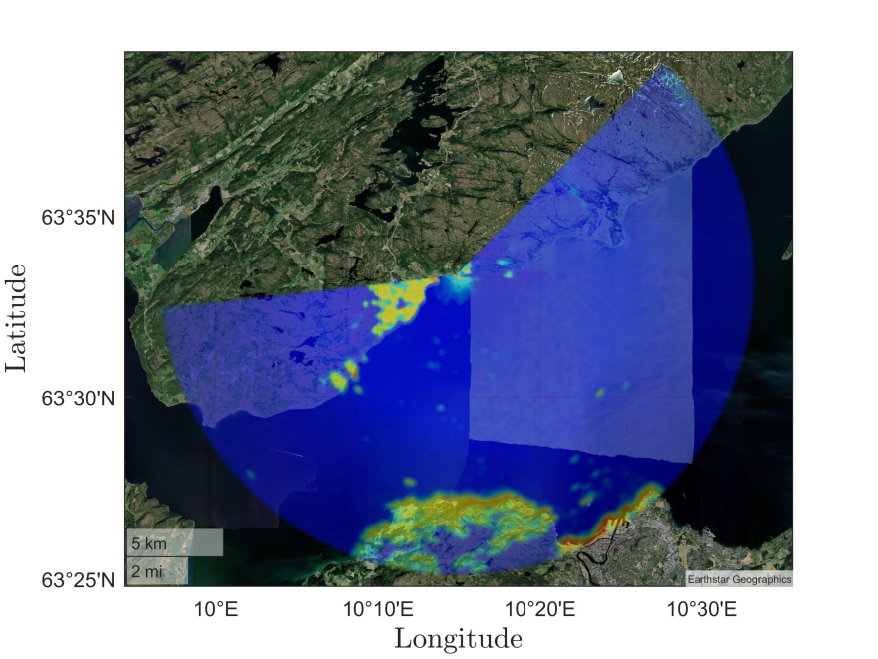}
    \caption{A single radar frame overlaid on a satellite image, showing the radar's location and the area of the Trondheim Fjord, Norway, observed by the radar system.}
    \label{fig:RadarFoV}
\end{figure}

The experimental system is based on a shore-based X-band marine surveillance radar located near Trondheim, Norway. Positioned at a slightly elevated coastal site, the radar provides high-resolution scans over a wide \textit{field of view} (FoV) covering a substantial area of the Trondheim Fjord, including ferry terminals, cruise and cargo shipping lanes, and harbour facilities. Fig.~\ref{fig:RadarFoV} shows the area observed by the radar.

This multi-target tracking estimation problem is further complicated by several real-world factors. First, the signal strength of targets can vary significantly due to factors such as low RCS, long ranges, multi-path fading, and partial occlusion. Second, sea clutter often exhibits heavier-tailed distributions and spatial or temporal correlation \cite{ward_sea_2013}. Third, the number of targets is not known in advance and may change over time. Additionally, targets can appear in close proximity to one another, resulting in overlapping returns that make discrimination more difficult.
These challenges motivate the use of TBD, which operates directly on the unthresholded measurements $Z_k$. However, TBD approaches are often computationally intensive, particularly in multi-target scenarios in large surveillance areas. Thus, a scalable tracking architecture is required that can combine the detection sensitivity of TBD with the efficiency of conventional tracking algorithms.

\begin{figure}[htb]
    \centering
    \includegraphics[width=0.995\linewidth]{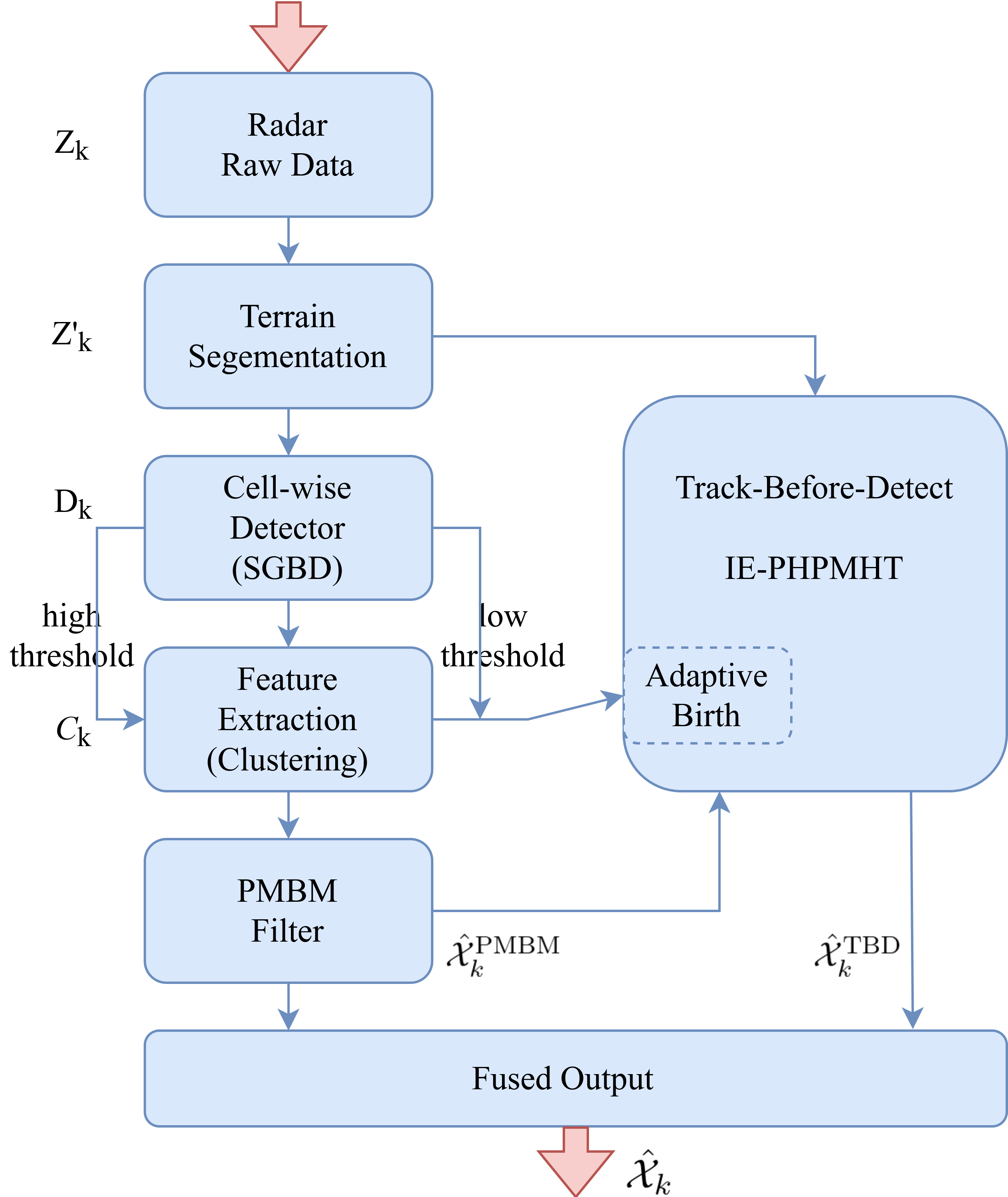}
    \caption{Block diagram of the full hybrid tracking architecture. The radar raw data first undergoes terrain segmentation, followed by a dual-threshold process applied by a cell-wise detector. In the first branch, high-threshold detections are clustered and converted into point observations, which are then used by the PMBM filter for recursive multi-target tracking. In the second branch, the IE-PHPMHT module performs TBD processing directly on the raw radar data, incorporating an adaptive birth mechanism and using feedback from the PMBM filter. The outputs are then fused for the multi-target state estimate.}
    \label{fig:SPChain}
\end{figure}

To address the tracking problem, we propose a \textit{hybrid tracking architecture} that integrates two complementary subsystems:
\begin{itemize}
    \item A PMBM filter, which processes high-threshold point detections. This subsystem aims to track high SNR targets.
    \item An IE-PHPMHT-based TBD tracker, which operates directly on the raw radar data. This subsystem aims to detect and track the low SNR targets.
\end{itemize}
This architecture enables the system to operate in real time while preserving the advantages of TBD for low-visibility targets and leveraging efficient point-tracking for high-confidence detections. 
The tracking system has the following modules:
\begin{enumerate}
    \item \textit{Terrain Segmentation:} Raw radar data is processed to remove persistent land clutter and stationary interference. 
    \item \textit{Detection and Feature Extraction:} A dual-threshold scheme is applied, i.e., high-threshold detections are clustered and converted into point observations for the PMBM filter, and low-threshold radar returns are extracted and given to the TBD module as potential target birth locations.
    \item \textit{Tracking Framework:} The PMBM filter performs recursive multi-target filtering based on the extracted point detections. The IE-PHPMHT performs recursive multi-target TBD working directly on raw radar data. Further, the estimates of the PMBM tracker are provided to the IE-PHPMHT to avoid redundancy and reduce computational complexity.
    \item \textit{Track Fusion and Output Generation}: The outputs from both tracking algorithms are combined to generate a consistent multi-target state estimate.
\end{enumerate}
The concept is visualised in the block diagram in Fig.~\ref{fig:SPChain} and the individual modules are detailed in the subsequent Section~\ref{sec:framework}.

%% file: 03_Framework.tex
\section{Hybrid Tracking Architecture}
\label{sec:framework}
This section outlines the individual components of the hybrid tracking architecture illustrated in Fig. \ref{fig:SPChain}. Section \ref{sec:arch_terrain} describes the terrain segmentation and land-clutter removal, followed by the detection process in Section \ref{sec:arch_detection} and clustering in Section \ref{sec:arch_clustering}. The two core tracking algorithms are then introduced and discussed in Sections \ref{sec:IE_PHPMHT} and \ref{sec:PMBM}, respectively, and the track fusion is explained in Section \ref{sec:arch_output}.

\subsection{Terrain Segmentation}
\label{sec:arch_terrain}
In radar-based maritime surveillance, static clutter from land has a significant impact on the tracker, as strong backscattered signals cause false detections. To mitigate this, a preprocessing step involves identifying and suppressing radar reflections originating from land surfaces (including other static objects such as buoys, dykes, etc.). In this work, we employ a precomputed median-based land mask generated offline from a large historical radar dataset and apply it at each time step to remove land clutter from the radar frames to be processed by the tracking algorithm.

Let the radar data cube be defined as $Z=(Z_1,\dots,Z_K)$, capturing $K$ time steps (frames), and each frame corresponds to a radar intensity image $Z_k$. To identify areas that consistently return strong radar reflections, we compute a background reference image $B = \left[b^1, \dots, b^i, \dots, b^I\right]^\transp$ using the temporal median over a large offline dataset. This is computed cell-wise as:
\begin{equation}
b^i = \underset{k \in \mathcal{K}_{\text{train}}}{\text{median}} \ z^i_k\,, \quad \forall i \in \{1, \dots, I\},
\end{equation}
where $\mathcal{K}_\text{train}$ is the set of radar frames used.
Subsequently, a binary land mask is calculated by thresholding the precomputed background:
\begin{equation}
\ell^i =
\begin{cases}
1, & \text{if } b^i > \tau, \\
0, & \text{otherwise},
\end{cases}
\quad \forall i \in \{1, \dots, I\},
\end{equation}
where $b^i$ is the background intensity in the $i$-th cell and $\tau \in \mathbb{R}$ is a manually selected threshold. This results in a binary mask $L = \left[\ell^1, \dots, \ell^i, \dots, \ell^I\right]^\transp \in \{0,1\}$ where $\ell^i = 1$ denotes a cell corresponding to land and and  $\ell^i = 0$ denotes a cell corresponding to sea.

Radar reflections from land often spread beyond their source. To ensure that all clutter associated with land is removed and not only the centre peaks, the mask $L$ is expanded using a morphological dilation operation. This involves enlarging each land pixel to cover a surrounding rectangular area, defined by a chosen number of pixels in both azimuth and range directions, leading to a dilated mask $L' = \left[\ell'^1, \dots, \ell'^i, \dots, \ell'^I\right]^\transp$.

At runtime, each incoming radar frame $Z_k$ is processed by applying the binary mask. Since $L'$ is static, it can be used for each new time step. The masked frame after land clutter removal is:
\begin{equation}
z'^i_k =
\begin{cases}
0, & \text{if } \ell'^i = 1, \\
z^i_k, & \text{if } \ell'^i = 0,
\end{cases}
\quad \forall i \in \{1, \dots, I\},
\end{equation}
resulting in the filtered frame $Z'_k = \left[z'^1_k, \dots, z'^i_k, \dots, z'^I_k\right]^\transp$.

\subsection{Cell-wise Detection}
\label{sec:arch_detection}
To obtain detections $D_k = \left[d_k^1, \dots, d_k^i, \dots, d_k^I\right]^\transp$ at time $k$ from a radar image $Z_k$, we utilise the \textit{spatial gradient-based detector} (SGBD) introduced in \cite{herrmann_coherent_2024,herrmann_target_2025}. This method leverages spatial image gradients to identify target-like structures through a non-parametric approach, without requiring prior knowledge of the target appearance or statistical clutter models. Instead of applying direct thresholding as in traditional \textit{constant false alarm rate} (CFAR) detectors~\cite{richards_fundamentals_2014}, the algorithm begins by computing the spatial gradients of $Z_k$, highlighting local variations indicative of object boundaries. These gradients are then regularised by enforcing a spatial smoothness constraint that reduces the influence of noise. The result is a vector field that aligns with structural features in the image while suppressing noise-induced fluctuations. From this smoothed gradient field, the divergence is computed, capturing regions where intensity changes rapidly and consistently, typically corresponding to target-like patterns. The divergence map is then thresholded to yield a binary map of candidate detections. The final output $D_k\in \{0,1\}$ is a binary detection mask indicating the locations of likely targets in the image $Z_k$. In \cite{herrmann_coherent_2024}, it has been shown that by exploiting gradient information rather than absolute intensity, SGBD enhances detection reliability for low-SNR targets, while maintaining a low false alarm rate under the heavy-tailed or correlated clutter conditions typical of maritime radar data, in comparison to standard CFAR methods.

\subsection{Feature Extraction (Clustering)}
\label{sec:arch_clustering}
The output of the SGBD is a binary map $D_k$, where each resolution cell is classified as either background or target-like.
Because radar returns from a single target typically span multiple neighbouring cells, the elements of $D_k$ tend to form spatially coherent clusters. To group these into meaningful detections, we apply a clustering step that aggregates detections likely originating from the same physical object. We use \textit{density-based spatial clustering of applications with noise} (DBSCAN), which groups spatially proximate detections without requiring prior knowledge of the number of targets~\cite{ester_density-based_1996}. As a result, a new set of detections $\mathcal{C}_k = \left\{c_k^1, \dots, c_k^{N_k^{\mathcal{C}}}\right\}$ is obtained, where $N_k^{\mathcal{C}}$ is the number of clusters at time frame $k$, and each $c_k$ represents the centre location of a cluster of resolution cells exceeding the SGBD detection threshold. These clustered detections are provided to the subsequent tracking stages, either as a direct input to the PMBM filter or as an input for the adaptive birth within the IE-PHPMHT.


\input{IE-PHPMHT}

\input{PMBM}

\subsection{Track Fusion and Output Generation}
\label{sec:arch_output}
The final stage of the hybrid tracking architecture combines the outputs from both the PMBM and the IE-PHPMHT tracking modules.


Within the IE-PHPMHT, a gating-based mechanism is applied during the adaptive birth phase. When a new potential birth state \( x_k^b \) is identified from low-threshold radar returns, it is compared against the current set of PMBM estimates \( \hat{\mathcal{X}}_k^{\text{PMBM}} \). If there exists at least one PMBM estimate \( \hat{x}_k^{\text{PMBM}} \in \hat{\mathcal{X}}_k^{\text{PMBM}} \) such that the Euclidean distance between the proposed birth position \( p_k^b \) and the estimated position \( \hat{p}_k^{\text{PMBM}} \) satisfies
$
\left\| p_k^b - \hat{p}_k^{\text{PMBM}} \right\| \leq \varepsilon,
$
no new Bernoulli component is added in the IE-PHPMHT. Here, \( p_k^b \) and \( \hat{p}_k^{\text{PMBM}} \) denote the position components subspace of the respective state vectors.
In this paper, a value of $\varepsilon = 100\,\text{m}$ is used, as a practical trade-off between preventing initiating redundant tracks and, to some extent, preserving sensitivity to spatially distinct, low-SNR targets.

Once both tracking modules have completed their respective filtering steps at time $k$, each provides estimated target states. The PMBM module outputs tracks based on high-threshold detections, while the IE-PHPMHT module contributes estimates $\hat{\mathcal{X}}_k^\text{TBD}$ derived from unthresholded data where the targets falling under the previous criteria are not instantiated. Hence, the final multi-target set estimate is formed by the union of these two outputs:
\begin{equation}
\hat{\mathcal{X}}_k = \hat{\mathcal{X}}^{\text{PMBM}}_k \cup \hat{\mathcal{X}}^{\text{TBD}}_k.
\end{equation}

%% file: IE-PHPMHT.tex
\subsection{Track-Before-Detect: IE-PHPMHT}
\label{sec:IE_PHPMHT}
We consider the IE-PHPMHT \cite{herrmann_histogram-probabilistic_2025}, with its main steps including the dynamic model, measurement and quantisation models, expectation-maximisation, and the computation of existence probabilities. We proceed to describe these steps. Details can be found in \cite{herrmann_histogram-probabilistic_2025}.


\subsubsection{Dynamic models}
The targets are assumed to evolve independently over time.
To capture target births and deaths, each potential target~$m$ is associated with a binary random variable $e_\timeind^{m} \in \{0,1\}$, with $e_\timeind^{m}=1$ indicating presence and $e_\timeind^{m}=0$ indicating absence at time~$\timeind$. The collection of all existence variables is denoted $E_\timeind = [\,e_\timeind^{1},\ldots,e_\timeind^{m},\ldots,e_\timeind^{\Targetind}\,]^{\transp}$.

New targets appear according to a multi-Bernoulli birth model. 
This model represents that there can be $M_\timeind^{b}$ targets born at time step $k$, each with a constant probability of birth $p_b$ and a single-target density $b(\cdot)$~\cite{mahler_advances_2014}.
Each existing target at $\timeind-1$ can either survive to the next time step with constant probability $p_s$ and transition to a new state or die with a probability of $1-p_s$. Conditional on survival, the kinematic state propagates via the known Markov transition density $ p\bigl(x_\timeind^{m}| x_{\timeind-1}^{m}\bigr) $.
The number of hypothesised targets after the prediction is $\Targetind = M_{\timeind-1}^{s}+M_\timeind^{b}$, where $M_{\timeind-1}^{s}$ is the number of survived targets. We assume that the states of different objects die or survive and evolve independently. 

\subsubsection{Measurement model}
The fundamental principle underlying the PHPMHT framework is that the observed measurements result from an underlying mixture process consisting of the aggregation of contributions from both targets and clutter. In the derivation, the continuous intensity values within each resolution cell \mathintext{\pixelind} are quantised, producing histogram data \mathintext{N_\timeind = \bigl[n_\timeind^{1}, \ldots, n_\timeind^{i}, \ldots, n_\timeind^{\Pixelind}\bigr]^{\transp}}, where every element contains a random variable representing the total number of measurements \mathintext{\makesubandsuper{n}{\timeind}{\pixelind}} that are located inside each cell \mathintext{i}, and \mathintext{\Pixelind} is the total number of resolution cells. The \textit{Poisson point process} (PPP) intensity function is expressed as:
\begin{align}
\label{eq:meas_mod_intensity}
\nu_\timeind\!\bigl(y | X_\timeind,\Lambda_\timeind\bigr)
    &= \lambda_\timeind^{0}\,p^{0}(y)
       +\sum_{m=1}^{\Targetind}\lambda_\timeind^{m}\,
         p^{m}\bigl(y | x_\timeind^{m}\bigr),  
\end{align}
where $y\!\in\!\mathbb{R}^{n_{y}}$ is a measurement location, $p^{0}(y)$ denotes the spatial density of clutter, and $p^{m}(y| x_\timeind^{m})$ denotes the known spatial influence of target~$m$. The vector of Poisson rates is $ \Lambda_\timeind = [\lambda_\timeind^{0},\ldots,\lambda_\timeind^{m},\ldots,\lambda_\timeind^{\Targetind}]^{\transp}$.
Integrating the intensity ~\eqref{eq:meas_mod_intensity} over the spatial extent~$A^{i}$ of resolution cell~$i$ yields the Poisson rate which depends on $X_k$ and $\Lambda_k$:
\begin{align}
\label{eq:meas_mod_rate}
\Bar{\nu}_\timeind^{i}\!\bigl(X_\timeind,\Lambda_\timeind\bigr)
    &= \lambda_\timeind^{0}\!\!\int_{A^{i}}\!p^{0}(y)\,\mathrm{d}y
       +\sum_{m=1}^{\Targetind}\lambda_\timeind^{m}\!\!
         \int_{A^{i}}\!p^{m}\bigl(y | x_\timeind^{m}\bigr)\,\mathrm{d}y.
\end{align}
Finally, because the superposition of independent PPPs is again Poisson, the total number of measurements in cell~$i$ follows:
\begin{align}
\label{eq:cardinality}
p\left(n_\timeind^{i}\right)
    &= \mathrm{e}^{-\Bar{\nu}_\timeind^{i}}\,
       \frac{\bigl(\Bar{\nu}_\timeind^{i}\bigr)^{\,n_\timeind^{i}}}{n_\timeind^{i}!}.
\end{align}

\subsubsection{Quantisation for track-before-detect}
In radar TBD, we make use of the entire sensor image where the sensor delivers continuous intensity imagery rather than a discrete histogram count. Therefore, the H-PMHT framework accommodates this by interpreting each intensity value as the limiting case of infinitesimal quantisation~\cite{streit_tracking_2000}.  
Let $ Z_\timeind = [\,z_\timeind^{1},\ldots,z_\timeind^{i},\ldots,z_\timeind^{\Pixelind}]^{\transp} $ denote the raw radar intensities. Introducing an arbitrary quantum~$\hbar^{2}$, we define $ n_\timeind^{i} = \bigl\lfloor z_\timeind^{i}/\hbar^{2}\bigr\rfloor = z_\timeind^{i}/\hbar^{2} - \epsilon$ with $\quad 0\le\epsilon<1$. Letting $\hbar^{2}\to 0$ yields~\cite{davey_track-before-detect_2018}:
\begin{align}
\label{eq:quant}
\lim_{\hbar^{2}\to 0}\hbar^{2}n_\timeind^{i}
    &= z_\timeind^{i},
&
\lim_{\hbar^{2}\to 0}\hbar^{2}\bigl\lVert N_\timeind\bigr\rVert_{1}
    &= \bigl\lVert Z_\timeind\bigr\rVert_{1} \;,
\end{align}
where $\lVert\cdot\rVert_{1}$ denots the $L_1$-norm.
Consequently, the quantisation is solely part of the modelling, but the algorithm can be applied directly to the input intensity data $Z_\timeind$ without explicit quantisation.

\subsubsection{Expectation maximisation}
From the measurement model \eqref{eq:meas_mod_rate}, it can be seen that the data observed by the sensor is incomplete, and we are concerned with two latent variables. Specifically, first, the component $\targetind$ from which the measurement $y$ originated, and second, the precise location of that measurement $y$ within the resolution cell $i$. To address this, two auxiliary variables $m_k \in \{0,1,\ldots,M_k\}$ and $y_k^i \in A^i$ are used such that the intensity of the PPP is:
\begin{align}
\label{eq:measurement_model_wo_hidden}
&\nu_\timeind^{\pixelind}\left(m_\timeind,\makesubandsuper{\vect{y}}{\timeind}{\pixelind}|X_\timeind,\Lambda_\timeind\right) \nonumber \\ 
& = \begin{cases}
\lambda_\timeind^{0}p^{0}(\makesubandsuper{\vect{y}}{\timeind}{\pixelind}) \mathds{1}_{A^\pixelind}(\makesubandsuper{\vect{y}}{\timeind}{\pixelind}) & m_\timeind=0\\
\lambda_\timeind^{m_\timeind}p^{m_\timeind}(\makesubandsuper{\vect{y}}{\timeind}{\pixelind}|\vect{x}_\timeind^{m_\timeind}) \mathds{1}_{A^\pixelind}(\makesubandsuper{\vect{y}}{\timeind}{\pixelind}) & m_\timeind\in\mathcal{M}_\timeind \;,
\end{cases}
\end{align}
where \mathintext{\mathds{1}_{A^\pixelind}(\makesubandsuper{\vect{y}}{\timeind}{\pixelind})} is the indicator function.
In the IE-PHPMHT, the missing data is a sequence of samples \mathintext{\mathbb{Y}_\timeind = \left( \makesubandsuper{\mathcal{Y}}{\timeind}{1},\ldots, \makesubandsuper{\mathcal{Y}}{\timeind}{\Pixelind} \right)} from this PPP, where each element is a set of points $ \makesubandsuper{\mathcal{Y}}{\timeind}{\pixelind}~=~\left\{ \left(m_{\timeind}^{\pixelind,1},\vect{y}_{\timeind}^{\pixelind,1}\right),...,\left(m_{\timeind}^{\pixelind,n_\timeind^{\pixelind}},\vect{y}_{\timeind}^{\pixelind,n_\timeind^{\pixelind}}\right)\right\}$.
Thus, the likelihood of the missing data is \cite{herrmann_histogram-probabilistic_2025}:
\begin{align}
\label{eq:lh_measurements}
    p(\mathbb{Y}_\timeind|N_\timeind,X_\timeind,\Lambda_\timeind) 
    &= \makesubandsuper{\prod}{\pixelindx=1}{\Pixelindx} \frac{n_\timeind^{\pixelind}!\prod_{r=1}^{n_\timeind^{\pixelind}} \makesubandsuper{\lambda}{\timeind}{\makesubandsuper{m}{\timeind}{\pixelind,r}} \makesubandsuper{p}{}{i,\makesubandsuper{m}{\timeind}{\pixelind,r}}\left( \makesubandsuper{\vect{y}}{\timeind}{\pixelind,r}|\makesubandsuper{\vect{x}}{\timeind}{\makesubandsuper{m}{\timeind}{\pixelind,r}} \right)}{\left(\Bar{\nu}_\timeind^{\pixelind}\right)^{n_\timeind^{\pixelind}}} .
\end{align}

with
\begin{equation}
    \makesubandsuper{p}{}{\makesubandsuper{i,m}{\timeind}{\pixelind,r}}\left( \makesubandsuper{\vect{y}}{\timeind}{\pixelind,r}|\makesubandsuper{\vect{x}}{\timeind}{\makesubandsuper{m}{\timeind}{\pixelind,r}} \right) = \makesubandsuper{p}{}{\makesubandsuper{m}{\timeind}{\pixelind,r}}\left( \makesubandsuper{\vect{y}}{\timeind}{\pixelind,r}|\makesubandsuper{\vect{x}}{\timeind}{\makesubandsuper{m}{\timeind}{\pixelind,r}} \right) \,
    \mathds{1}_{A^\pixelind}\left(\makesubandsuper{\vect{y}}{\timeind}{\pixelind,r}\right)\,.
\end{equation}

\textit{Expectation maximisation} (EM) \cite{bishop_pattern_2006} is employed to obtain maximum a-posteriori (MAP) estimates of the target states and their associated Poisson measurement rates from observed data, while marginalising out the unobserved latent variables. Here, the observed data consist of cell-wise measurement counts \( N_\timeind \), while the parameters of interest are the target states \( X_\timeind \) and the Poisson rates \( \Lambda_\timeind \). Instead of directly maximising the posterior distribution, EM proceeds by iteratively constructing an auxiliary function $\mathcal{L}$ that is maximised at each iteration \( \itind \):
\begin{align}
    \left(\hat{X}_\timeind^{(\itind)},\hat{\Lambda}_\timeind^{(\itind)}\right)_{\text{MAP}} &= \underset{X_\timeind,\Lambda_\timeind}{\text{argmax}} \ \makesubandsuper{\aux}{\timeind}{(\itind)}\left(X_\timeind,\Lambda_\timeind|\hat{X}_\timeind^{(\itind-1)},\hat{\Lambda}_\timeind^{(\itind-1)} \right)\;.
\end{align}
This function is defined as the expected log-likelihood of the complete data, comprising both observed and latent components, conditioned on the parameter estimates from the previous iteration:
\begin{align}
\label{eq:aux}
    &\mathcal{L}_\timeind^{(\itind)}(X_\timeind, \Lambda_\timeind | \hat{X}_\timeind^{(\itind-1)}, \hat{\Lambda}_\timeind^{(\itind-1)}) \nonumber\\
    &= \mathbb{E}_{\mathbb{Y}_\timeind | N_\timeind, \hat{X}_\timeind^{(\itind-1)}, \hat{\Lambda}_\timeind^{(\itind-1)}} 
\left[\log \left\{ p_{\text{c}}(N_\timeind, \mathbb{Y}_\timeind, X_\timeind, \Lambda_\timeind) \right\}\right] \nonumber\\
    &= \int \log  \left\{ p_{\text{c}}(N_\timeind, \mathbb{Y}_\timeind, X_\timeind, \Lambda_\timeind) \right\}
    \, \nonumber \\
    & \ \ \ \ \times p_{\text{m}}(\mathbb{Y}_\timeind | N_\timeind, \hat{X}_\timeind^{(\itind-1)}, \hat{\Lambda}_\timeind^{(\itind-1)}) 
    \, \delta \mathbb{Y}_\timeind \;,
\end{align}
where \( p_{\text{c}} \) denotes the complete-data likelihood, and \( \mathbb{Y}_\timeind \) is the sequence of sets representing the latent variables, and is therefore integrated via a set integral \cite{mahler_advances_2014}. The expectation is taken with respect to \( p_{\text{m}} \), which denotes the distribution of the latent variables conditioned on the observations and the previous parameter estimates.

For the expectation step, the explicit expression of the auxiliary function is constructed.
The complete data likelihood $p_c$ is factorised as:
\begin{align}
p_{\text{c}}(N_\timeind, \mathbb{Y}_\timeind, X_\timeind, \Lambda_\timeind) = \, &p(X_\timeind) \, p(\Lambda_\timeind) \, p(N_\timeind | X_\timeind, \Lambda_\timeind) \nonumber\\
\times &p(\mathbb{Y}_\timeind | N_\timeind, X_\timeind, \Lambda_\timeind),
\end{align}
where the components are the state prior $ p(X_\timeind)$, the Poisson rate prior $p(\Lambda_\timeind)$, the likelihood of the observed measurement counts $p(N_\timeind|X_\timeind, \Lambda_\timeind)$ given in \eqref{eq:cardinality} and the product over all resolution cells of the likelihood of the missing data $p(\mathbb{Y}_\timeind | N_\timeind, X_\timeind, \Lambda_\timeind)$ given in \eqref{eq:lh_measurements}.

The auxiliary function can be decomposed into two parts
$
\mathcal{L}_\timeind^{(\itind)} = \mathcal{L}_{\timeind, \Lambda}^{(\itind)} + \mathcal{L}_{\timeind, X}^{(\itind)}
$.
The part of the auxiliary function corresponding to the Poisson intensities is given by \cite{herrmann_histogram-probabilistic_2025}:
\begin{align}
\label{eq:aux_lambda_final}
\mathcal{L}_{\timeind, \Lambda}^{(\itind)} 
= \sum_{\targetind=0}^{\Targetind} \left[ 
\log \left\{ p(\lambda_\timeind^{\targetind}) \right\} 
+ \log \left\{ \text{e}^{-\lambda_\timeind^{\targetind}} (\lambda_\timeind^{\targetind})^{\bar{n}_\timeind^{\targetind}} \right\}
\right],
\end{align}
where the expected measurement count \( \bar{n}_\timeind^{\targetind} \) for each component is:
\begin{align}
\bar{n}_\timeind^{\targetind} = \sum_{i=1}^{\Pixelind} n_\timeind^i 
\cdot \frac{ \hat{\lambda}_\timeind^{\targetind, (\itind-1)} 
\int_{A^i} p^{i,\targetind}(y|\hat{\vect{x}}_\timeind^{\targetind, (\itind-1)}) dy}
{\sum_{s=0}^{\Targetind} \hat{\lambda}_\timeind^{s, (\itind-1)} 
\int_{A^i} p^{i,s}(y|\hat{\vect{x}}_\timeind^{s, (\itind-1)}) dy} \;.
\end{align}
Similarly, the component of the auxiliary function related to the target states is \cite{herrmann_histogram-probabilistic_2025}:
%
%
\begin{align}
\label{eq:aux_state_final}
\mathcal{L}_{\timeind, X}^{(\itind)} 
=& \sum_{\targetind=1}^{\Targetind} \Biggl[ 
 \rVert N\rVert_{1} \log \left\{ p(\vect{x}_\timeind^{\targetind}) \right\} 
+ \hat{\lambda}_\timeind^{\targetind,(\itind-1)} 
\sum_{i=1}^{\Pixelind} \frac{n_\timeind^i}{\hat{\bar{\nu}}_\timeind^{i,(\itind-1)}} \nonumber\\
& \times \int_{A^i} \log \left\{ p^{i,\targetind}(y|\vect{x}_\timeind^{\targetind}) \right\}
\, p^{i,\targetind}(y|\hat{\vect{x}}_\timeind^{\targetind,(\itind-1)}) \, dy 
\Biggr],
\end{align}
where \( \hat{\bar{\nu}}_\timeind^{i,(\itind-1)} \) is the total intensity in cell \( i \) computed using the previous estimates.

The maximisation step of the EM algorithm updates the model parameters by maximising the auxiliary function obtained during the expectation step.
The auxiliary function $\mathcal{L}_{\timeind,X}^{\itind}$ evaluates the expected log-likelihood of measurements generated under the model assumption of a PPP.
This can be interpreted as a weighted log-likelihood problem using synthetic measurements \cite{streit_tracking_2000}. For linear Gaussian models i.e., $p^\targetind(y|\makesubandsuper{x}{\timeind}{\targetind}) = \Gaussian(y; H\makesubandsuper{x}{\timeind}{\targetind}, R)$, the optimal solution corresponds to applying a Kalman filter (KF) using the synthetic measurements \cite{streit_tracking_2000}.
The synthetic measurement for target \(\targetind\) at time step \(\timeind\) is defined using the raw intensity data \(Z_\timeind\), replacing the quantised count data using \eqref{eq:quant}. The synthetic measurement is then computed as \cite{streit_tracking_2000}:
\begin{align}
\label{eq:synthetic_measurement_intensity}
    \makesubandsuper{\tilde{z}}{\timeind}{\targetind} 
    = \frac{1}{\makesubandsuper{\hat{\Bar{\nu}}}{\timeind}{\targetind}} 
    \sum_{\pixelind=1}^{\Pixelindx} z_\timeind^\pixelind 
    \int_{A^\pixelind} y \, \makesubandsuper{\hat{\lambda}}{\timeind}{\targetind} \, 
    p^{i,\targetind}(y|\makesubandsuper{\hat{x}}{\timeind}{\targetind}) \, dy \;,
\end{align}
where \(z_\timeind^\pixelind\) is the intensity measured in resolution cell \(\pixelind\), and \(\makesubandsuper{\hat{\Bar{\nu}}}{\timeind}{\targetind}\) denotes the expected total contribution from target \(\targetind\), given by $
\label{eq:expected_measurements_target_intensity}
    \makesubandsuper{\hat{\Bar{\nu}}}{\timeind}{\targetind} 
    = \sum_{\pixelind=1}^{\Pixelindx} \makesubandsuper{\hat{\lambda}}{\timeind}{\targetind} 
    \int_{A^\pixelind} p^{i,\targetind}(y|\makesubandsuper{\hat{x}}{\timeind}{\targetind}) \, dy
$.
The synthetic covariance is given as \cite{streit_tracking_2000}:
\begin{align}
\label{eq:synthetic_covariance_intensity}
    \makesubandsuper{\tilde{R}}{\timeind}{\targetind} 
    = R \left( 
    \sum_{\pixelind=1}^{\Pixelindx} 
    \frac{z_\timeind^\pixelind \, \makesubandsuper{\hat{\lambda}}{\timeind}{\targetind} 
    \int_{A^\pixelind} p^{i,\targetind}(y|\makesubandsuper{\hat{x}}{\timeind}{\targetind}) \, dy}{
    \makesubandsuper{\hat{\Bar{\nu}}}{\timeind}{\pixelind}} 
    \right)^{-1} \;,
\end{align}
where $R$ models the target spread, and \(\makesubandsuper{\hat{\Bar{\nu}}}{\timeind}{\pixelind}\) is the estimated total intensity in resolution cell \(\pixelind\).

The rate update maximises the auxiliary function $\mathcal{L}_{\timeind,\Lambda}^{\itind}$. With a Gamma prior $\mathcal{G}(\lambda_{\timeind}^{\targetind}; \alpha_{\timeind}^{\targetind}, \beta_{\timeind}^{\targetind})$ and Poisson likelihood (see \eqref{eq:aux_lambda_final}), the posterior is also Gamma-distributed:
$
    p(\lambda_{\timeind}^{\targetind} | \bar{n}_{\timeind}^{\targetind}) 
    = \mathcal{G}(\lambda_{\timeind}^{\targetind}; a_{\timeind}^{\targetind}, b_{\timeind}^{\targetind}),
$
with updated shape and rate parameters:
$ a_{\timeind}^{\targetind} = \alpha_{\timeind}^{\targetind} + \bar{n}_{\timeind}^{\targetind}$ and $b_{\timeind}^{\targetind} = \beta_{\timeind}^{\targetind} + 1$.
The MAP estimate for the Poisson rate is thus given by the mode of the posterior:
\begin{align}
    \hat{\lambda}_{\timeind}^{\targetind} = \frac{a_{\timeind}^{\targetind} - 1}{b_{\timeind}^{\targetind}}.
\end{align}
These updates are performed iteratively until convergence of the auxiliary function or until a fixed number of EM iterations is reached.

\subsubsection{Recovering the probability of existence}

After the EM update, we obtain point estimates of the Poisson rates $\makesubandsuper{\hat{\lambda}}{\timeind}{\targetind}$ for each potential target. To determine whether a target likely exists, we compute the posterior probability of existence conditioned on the Poisson rate $\makesubandsuper{r}{\timeind}{\targetind} = p(e_\timeind^{\targetind} = 1 | \lambda_\timeind^{\targetind})$.
It is assumed that the Poisson rate $\lambda_\timeind^{\targetind}$ depends on whether the target exists $e_\timeind^{\targetind} = 1$, and the Poisson rate prior is modelled with a Gamma distribution $p(\lambda_\timeind^{\targetind}| e_\timeind^{\targetind}=1) = \mathcal{G}(\lambda_\timeind^{\targetind}; \alpha_\timeind^{\targetind}, \beta_\timeind^{\targetind})$, or the target does not exist $e_\timeind^{\targetind} = 0$ and the prior of the Poisson rate is an exponential distribution $p(\lambda_\timeind^{\targetind}| e_\timeind^{\targetind}=0) = \mathcal{E}(\lambda_\timeind^{\targetind}; \gamma_\timeind^{\targetind})$. Using Bayes' rule, the probability of existence can be calculated as:
\begin{align}
\label{eq:calc_prob_ex_final}
    \makesubandsuper{r}{\timeind}{\targetind} =
    \left.
    \frac{
        \mathcal{G}(\lambda; \alpha_\timeind^{\targetind}, \beta_\timeind^{\targetind}) \cdot r_{\timeind|\timeind-1}^{\targetind}
    }{
        \mathcal{G}(\lambda; \alpha_\timeind^{\targetind}, \beta_\timeind^{\targetind}) \cdot r_{\timeind|\timeind-1}^{\targetind}
        +
        \mathcal{E}(\lambda; \gamma_\timeind^{\targetind}) \cdot (1 - r_{\timeind|\timeind-1}^{\targetind})
    }
    \right|_{\lambda = \hat{\lambda}_\timeind^{\targetind}},
\end{align}
evaluated at the $\hat{\lambda}_\timeind^{\targetind}$ that results from EM and where $r_{\timeind|\timeind-1}^{\targetind}$ is the predicted probability of existence.

The EM update assumes a conjugate prior for the Poisson rate to ensure analytical tractability. However, due to the marginalisation of the existence variable $\sum_{\makesubandsuper{e}{\timeind}{\targetind}} p(\makesubandsuper{\lambda}{\timeind}{\targetind}|\makesubandsuper{e}{\timeind}{\targetind})p(\makesubandsuper{e}{\timeind}{\targetind})$, the prior for $\lambda_\timeind^{\targetind}$ becomes a mixture of a gamma and an exponential distribution \cite{herrmann_histogram-probabilistic_2025}:
\begin{align}
\label{eq:prior_lambda_existence}
    p(\lambda_\timeind^{\targetind}) =
    r_{\timeind|\timeind-1}^{\targetind} \cdot \mathcal{G}(\lambda_\timeind^{\targetind}; \alpha_\timeind^{\targetind}, \beta_\timeind^{\targetind})
    +
    (1 - r_{\timeind|\timeind-1}^{\targetind}) \cdot \mathcal{E}(\lambda_\timeind^{\targetind}; \gamma_\timeind^{\targetind}) .
\end{align}
To retain a closed-form EM update, the mixture is approximated with a single Gamma distribution that minimises the Kullback–Leibler divergence (KLD) \cite{bishop_pattern_2006} to the exact prior $\mathcal{G}(\lambda; a_\timeind^{\targetind}, b_\timeind^{\targetind}) \approx p(\lambda_\timeind^{\targetind})$. The mixture merging is not part of the EM and is performed between the prediction and update.

%% file: PMBM.tex
\subsection{The Poisson Multi-Bernoulli Mixture Filter}
\label{sec:PMBM}
The PMBM filter \cite{williams_marginal_2015,garcia-fernandez_poisson_2018} is a state-of-the-art approach to multi-target tracking that provides a Bayesian solution for detection-based measurement models. It is based on RFSs, but its connections to joint probabilistic data association and multiple hypothesis tracking have been formally established in \cite{williams_marginal_2015,brekke_relationship_2018}. We proceed to describe the main characteristics of the PMBM filter. Details can be found in \cite{williams_marginal_2015,garcia-fernandez_poisson_2018}.

The problem of tracking multiple targets with both an unknown number of targets and an unknown number of measurements is addressed by the PMBM using RFSs.
The set of target states at time $k$ is represented by $\mathcal{X}_k=\left\{ x_k^1,\dots,x_k^{M_k} \right\}$ and the measurements obtained at time step $k$ as $\mathcal{C}_k = \left\{ c_k^1,\dots,c_k^{N_k^{\mathcal{C}}} \right\}$ with $c_k$ denoting a single measurement.

Moreover, the PMBM filter for point targets makes the following model assumptions. New targets appear via a PPP with intensity $\mu_k^b$. Each existing target at time step $ x_{k-1}$ survives with probability $p_s$ and transitions with density $p(x_k | x_{k-1})$ or dies with probability $1-p_s$, independently of other targets. A target is detected with probability $p_d$ and, if detected, generates a measurement via the likelihood $l(c | x_k)$. Clutter follows a PPP with intensity $\mu^\text{c}$. Measurements are assumed to originate from at most one object. To control the combinatorial growth in hypotheses over time, the PMBM filter usually applies hypothesis management techniques such as gating and pruning. These mechanisms ensure computational tractability and are detailed in Section~\ref{sec:Imp_Eval}.

The posterior PMBM density $f_{k'|k}(\cdot)$ represents the set of targets at time step $k'$ given the history of measurements up to time step $k$, and is a combination of a PPP and an MBM. The PPP, with density $f_{k'|k}^{\text{P}}(\cdot)$, represents the undetected targets $\mathcal{X}_{k'}^{\text{u}}$, and the MBM, with density $f_{k'|k}^{\text{MBM}}(\cdot)$, represents the detected targets $\mathcal{X}_{k'}^{\text{d}}$. The index $k'$ can be either $k$ or $k+1$, corresponding to the update and prediction steps, respectively.
The PMBM density, given measurements up to time $k$, is given by \cite{garcia-fernandez_poisson_2018}:
\begin{equation}
f_{k'|k}(\mathcal{X}_{k'}) = \sum_{\mathcal{X}_{k'}^{\text{u}} \uplus \mathcal{X}_{k'}^{\text{d}} = \mathcal{X}_{k'}} f^{\text{P}}_{k'|k}(\mathcal{X}_{k'}^{\text{u}}) \, f^{\text{MBM}}_{k'|k}(\mathcal{X}_{k'}^{\text{d}}),
\end{equation}
where $\mathcal{X}_{k'}^{\text{u}}$ and $\mathcal{X}_{k'}^{\text{d}}$ are mutually disjoint subsets whose union is $\mathcal{X_{k'}}$ corresponding to undetected and detected targets, respectively.
The PPP component is specified as \cite{garcia-fernandez_poisson_2018}:
\begin{equation}
f^{\text{P}}_{k'|k}(\mathcal{X}_{k'}^{\text{u}}) = e^{-\int \mu_{k'|k}(x) \, dx} \prod_{x \in \mathcal{X}_{k'}^{\text{u}}} \mu_{k'|k}(x),
\end{equation}
with $\mu_{k'|k}(x)$ being the intensity function of the PPP.
The MBM component of the PMBM density is given by \cite{garcia-fernandez_poisson_2018}:
\begin{equation}
f^{\text{MBM}}_{k'|k}(\mathcal{X}_{k'}^{\text{d}}) = 
\sum_{j \in \mathbb{J}_{k'|k}} w^{(j)}_{k'|k}
\sum_{\biguplus_{l \in \mathbb{I}_{k'|k}} \mathcal{X}_{k'}^l = \mathcal{X}_{k'}^{\text{d}}}
\prod_{i \in \mathbb{I}_{k'|k}} f^{(j,i)}_{k'|k}(\mathcal{X}_{k'}^i),
\end{equation}
where $\mathcal{X}_{k'}^{\text{d}}$ denotes the set of detected targets at time step $k'$, and $\mathbb{J}_{k'|k}$ is the index set of global data association hypotheses. Each global hypothesis $j \in \mathbb{J}_{k'|k}$ has an associated weight $w^{(j)}_{k'|k}$, satisfying $\sum_{j \in \mathbb{J}_{k'|k}} w^{(j)}_{k'|k} = 1$. The set $\mathbb{I}_{k'|k}$ indexes the individual Bernoulli components under each global hypothesis. The set $\mathcal{X}_{k'}^{\text{d}}$ is partitioned into disjoint subsets $\{\mathcal{X}_{k'}^i\}_{i \in \mathbb{I}_{k'|k}}$, such that $\biguplus_{l \in \mathbb{I}_{k'|k}} \mathcal{X}_{k'}^l = \mathcal{X}_{k'}^{\text{d}}$. Each subset $\mathcal{X}_{k'}^i$ is associated with a Bernoulli component whose density under hypothesis $j$ is denoted by $f^{(j,i)}_{k'|k}(\mathcal{X}_{k'}^i)$. The product of these Bernoulli densities defines a multi-Bernoulli distribution for each global hypothesis, and the full MBM density is obtained as a weighted mixture over all such hypotheses. The $j$-th multi-Bernoulli component has the probability $w_k^{(j)}$ and $N_k^{\mathcal{C}} = |\mathbb{I}_{k'|k}|$ Bernoulli components.

Each Bernoulli component under the global hypothesis $j$ has the form \cite{mahler_advances_2014}:
\begin{equation}
f^{(j,i)}_{k'|k}(\mathcal{X}_{k'}) =
\begin{cases}
1 - r^{(j,i)}_{k'|k} & \text{if } \mathcal{X}_{k'} = \emptyset, \\
r^{(j,i)}_{k'|k} \, p^{(j,i)}_{k'|k}(x_{k'}) & \text{if } \mathcal{X}_{k'} = \{x_{k'}\}, \\
0 & \text{otherwise},
\end{cases}
\end{equation}
where $r^{(j,i)}_{k'|k} \in [0,1]$ is the existence probability, and $p^{(j,i)}_{k'|k}(x_{k'})$ is the single-target spatial density.
In the Gaussian PMBM implementation \cite{garcia-fernandez_poisson_2018}, the single-target state density is Gaussian i.e., $ p^{(j,i)}_{k'|k}(x_{k'}) = \mathcal{N}(x_{k'}; \bar{x}^{(j,i)}_{k'|k}, P^{(j,i)}_{k'|k})$, where $\bar{x}^{(j,i)}_{k'|k}$ is the mean and $P^{(j,i)}_{k'|k}$ is the covariance matrix.
Hence, the full PMBM density $f_{k'|k}(\mathcal{X}_{k'})$ is parametrised by the following parameters
\begin{equation}
\mu_{k'|k}(x),
\left\{
w^{(j)}_{k'|k}, 
\left\{
r^{(j,i)}_{k'|k}, \bar{x}^{(j,i)}_{k'|k}, P^{(j,i)}_{k'|k}
\right\}_{i \in \mathbb{I}_{k'|k}}
\right\}_{j \in \mathbb{J}_{k'|k}}
\,,
\end{equation}
and is recursively estimated using the prediction and update step of the multi-target Bayes filter \cite{mahler_advances_2014}, and the recursions for prediction and update can be found in \cite{garcia-fernandez_poisson_2018}.

The PMBM filter has also been extended to extended object tracking \cite{granstrom_poisson_2020} with applications in the maritime domain \cite{baerveldt_multiple_2024} and sets of trajectories \cite{granstrom_poisson_2025}.

%% file: 04_Imp_ExpEval.tex
\section{Full-Scale Implementation and Experimental Evaluation}
\label{sec:Imp_Eval}
In this section, we present the full-scale implementation of the signal processing chain in Section \ref{sec:framework} and evaluate the performance based on experimental radar data. Section \ref{sec:Eval_setup} describes the experimental setup, while Section \ref{sec:Eval_impNotes} details the implementation of the different algorithms, and Section \ref{sec:Eval_results} presents the results of the experimental evaluation.

\subsection{Experimental Setup}
\label{sec:Eval_setup}
For the experimental evaluation, an X-band marine radar observing the Trondheim Fjord in Norway is employed (refer to Fig.\ref{fig:RadarFoV} for the system setup and the radar’s FoV). This radar is a stationary, shore-mounted, solid-state pulse-compression system that transmits linearly chirped pulses with its antenna mounted on a rotor. The radar specifications are listed in Table~\ref{tab:Radar_param}.

\begin{table}[htb]
\caption{X-Band Marine Radar Specification}
\label{tab:Radar_param}
\setlength{\tabcolsep}{3pt}
\renewcommand{\arraystretch}{1.3}  
\begin{tabular}{p{115pt}p{125pt}}
\hline\hline
\textbf{Parameter} & 
\textbf{Specification} \\
\hline
Frequency & 9.41\,GHz \\
Bandwidth & 5.8\,MHz \\
Antenna type \& length & Rotating slotted waveguide (8\,ft.) \\
Angular aperture azimuth (3\,dB) & $\approx$ 1\,$^\circ$ \\
Angular aperture elevation (3\,dB) & $\approx$ 20\,$^\circ$ \\
Polarisation & Horizontal \\
Gain & 32\,dBi \\
Transmitted power & 600\,W (peak) / $\approx$ 12\,W (avg.) \\
Antenna rotation speed & 24 RPM \\
Pulse repetition frequency & 2.5\,kHz \\
Number of range bins & 2048 \\
Number of azimuth bins & 720 \\
Detection range & 13.32\,km \\
Angular sector & 220\,$^\circ$ \\
Radar height (above sea level)& $\approx$ 15\,m \\
\hline
\end{tabular}
\end{table}

\begin{figure}[b!]
    \centering
    \begin{minipage}[t]{0.475\linewidth}
        \centering
        \includegraphics[width=\linewidth, height=2.5cm]{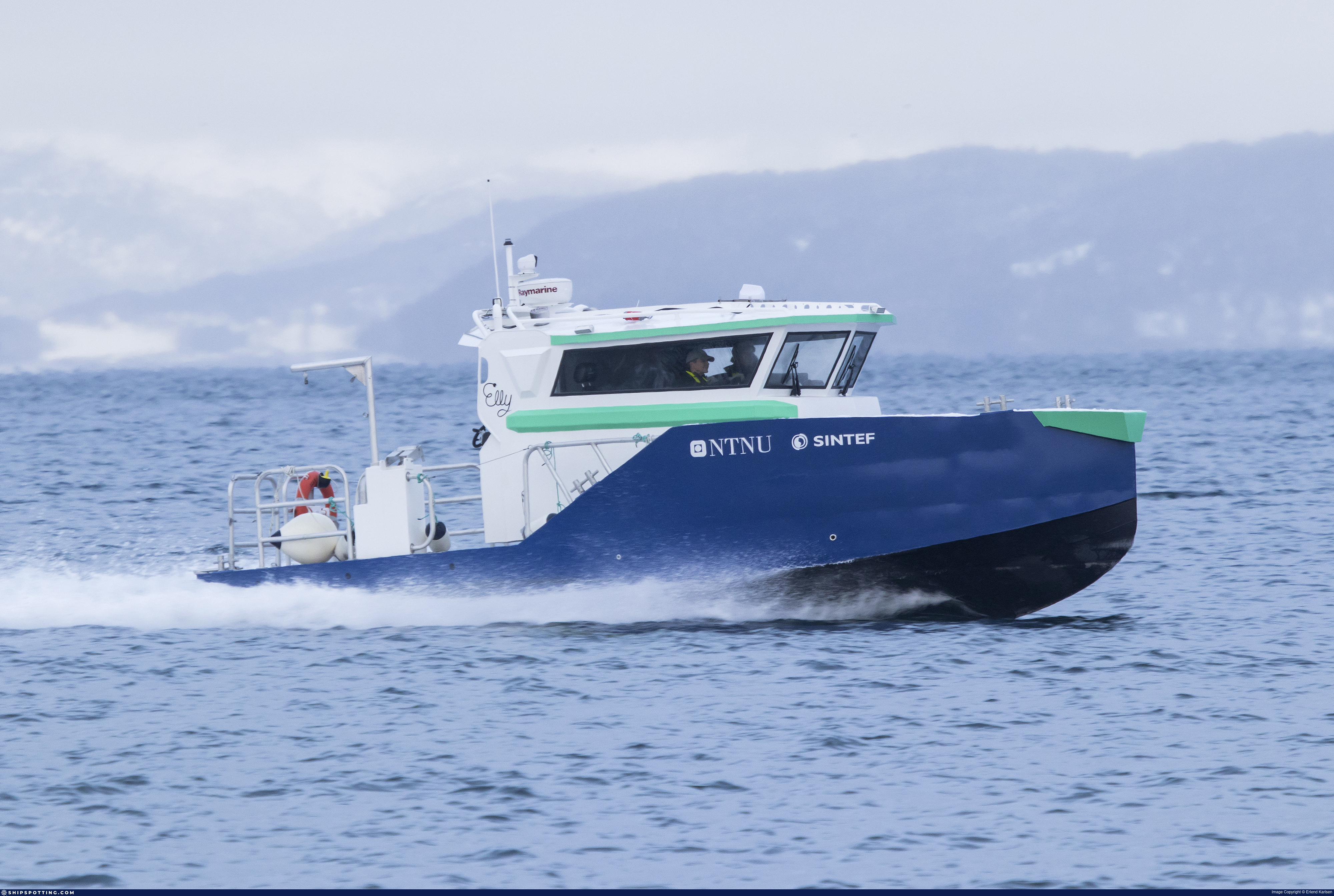}
        \footnotesize{(a)}
        \label{fig:elly}
    \end{minipage}
    \hfill
    \begin{minipage}[t]{0.475\linewidth}
        \centering
        \includegraphics[width=\linewidth, height=2.5cm]{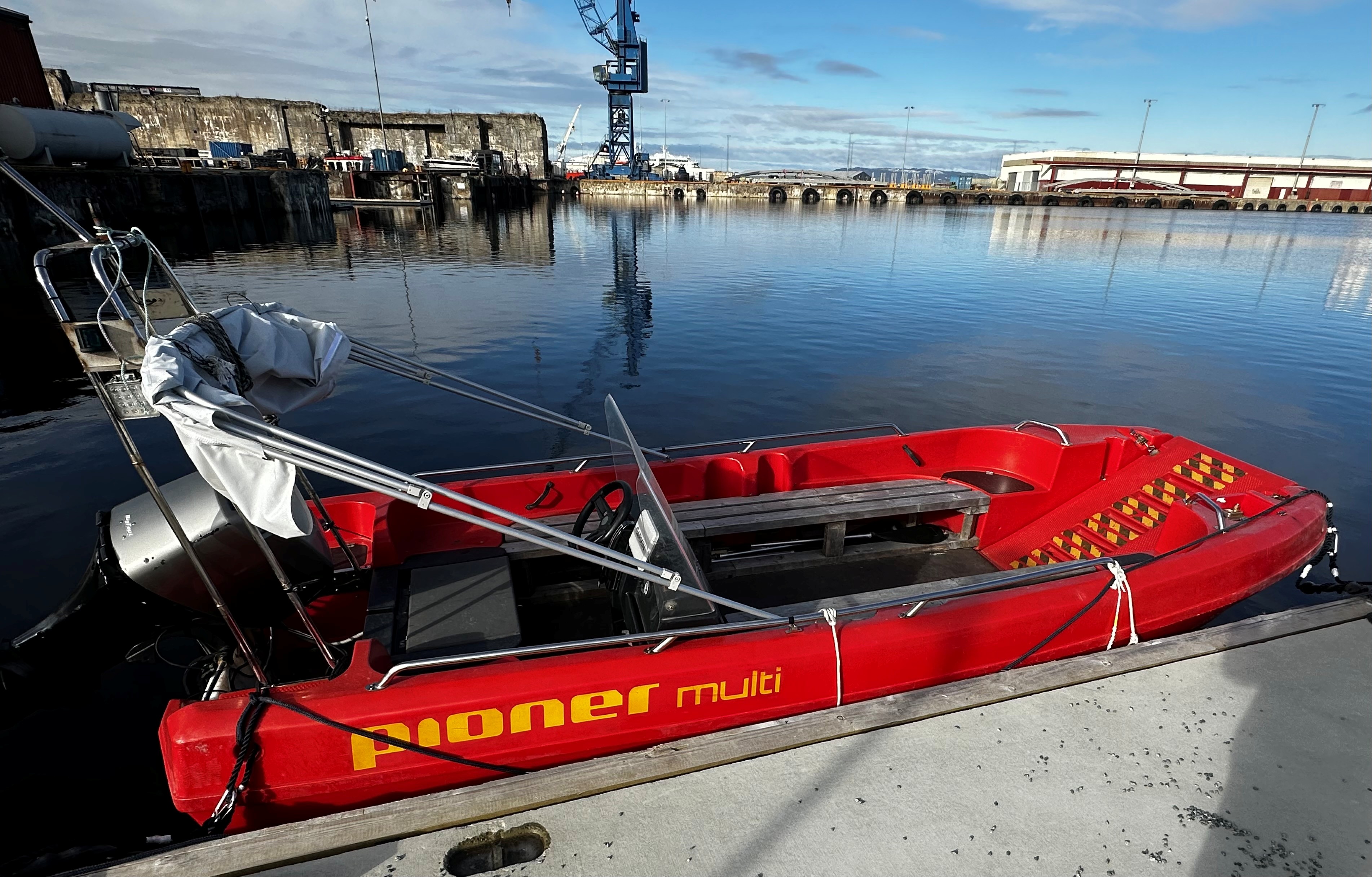}
        \footnotesize{(b)}
        \label{fig:pioner}
    \end{minipage}
    
    \caption{\footnotesize The two target vessels used in the experimental validation: {Elly} (a) and {Pioner} (b).}
    \label{fig:target_vessels}
\end{figure}
For the experiment, we used two target vessels: \say{Elly} (Fig. \ref{fig:elly}) and \say{Pioner} (Fig. \ref{fig:pioner}). The Pioner was deliberately selected for its low RCS. This is primarily due to its compact size (approximately 4.7\,m in length), low-profile hull that sits close to the sea surface, and construction from dielectric materials. While an exact RCS value is difficult to determine, the dominant contributors to the boat's RCS are essentially the two onboard operators and the outboard engine.
\begin{figure}[tb]
    \centering
    \includegraphics[width=\linewidth]{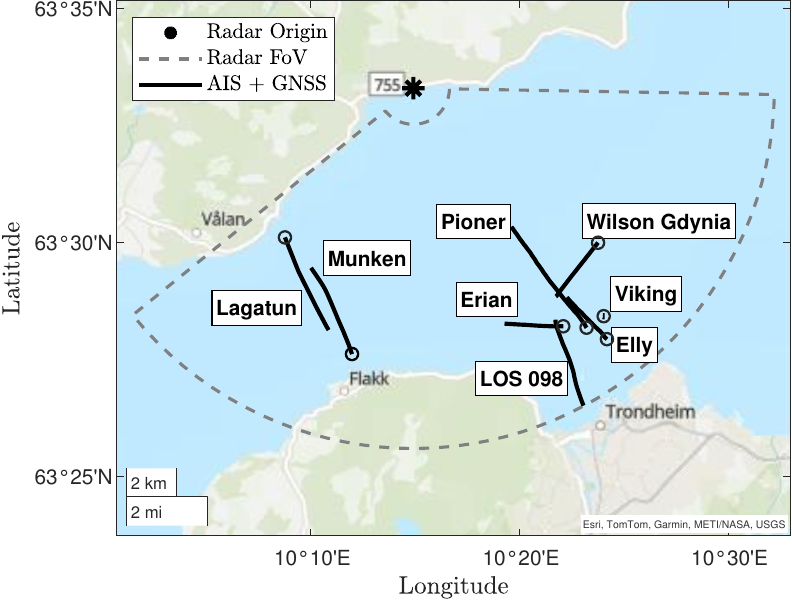}
    \caption{Reported AIS and GNSS ground truth trajectories during the experiment. The circles mark the starting positions.}
    \label{fig:Traj_GT_labeled}
\end{figure}
In Fig. \ref{fig:Traj_GT_labeled}, the trajectories of these two vessels, and other vessels during the experiments, are shown. 
During the experiment, two sources of ground truth data were recorded. First, the two target vessels operated by us were equipped with GNSS transceivers. Second, AIS\footnote{The use of AIS is widespread and mandated by international regulations for all passenger ships and for cargo ships of 300 gross tonnage or more when engaged on international voyages (https://www.imo.org/en/ourwork/safety/pages/AIS.aspx).} data were recorded for all the vessels present within the observed area during the experiment. Ground truth is used under the assumption that the reported positions are both accurate and error-free. It is further assumed that all vessels represented in the ground truth data are the only ones present within the surveillance area. Moreover, due to variability in the logging intervals of the AIS and GNSS data, which do not align with the radar frame updates, linear interpolation is applied to synchronise the ground truth positions to the radar time steps.

In the experiment, the two target vessels began their transit from the south side of the fjord, heading northwest toward the radar position. In addition to the target vessels, several other ships were present in the observed area: the fishing boat \say{Viking} was stationary positioned; a sailing boat \say{Erian} sailing from east to west; the car ferries \say{Munken} and \say{Lagatun} were crossing the fjord on the western side of Trondheim, operating on the Flakk–Rørvik route; the cargo vessel \say{Wilson Gdynia} was moving westward from the eastern fjord part; the pilot vessel \say{LOS 098}, departing from the Trondheim harbour, entered the area while approaching Wilson Gdynia.
The wind speed was approximately 4-6 m/s coming from the south-south-east.

The data were collected on the 21st of March 2025, and the collected data are given as a sequence of 250 raw radar range-azimuth maps, comprising 2048 range cells covering 13.32\,km and 720 azimuth resolution cells in a sector of 220 degrees. In this paper, we have processed range cells from 190 to 1900 and azimuth cells from 100 to 610, after the terrain segmentation. The resulting FoV is shown in e.g., Fig. \ref{fig:Traj_GT_labeled}. 

\subsection{Implementation Aspects}
\label{sec:Eval_impNotes}
In the following experimental results, the proposed hybrid TBD tracking algorithm, denoted as PMBM+IE-PHPMHT, is applied to the experimental data and analysed and compared with two variations of a point-based \textit{joint integrated probabilistic data association} (JIDPA) tracker \cite{musicki_joint_2004}. The used parameters of the PMBM, IE-PHPMHT, and JIPDA are presented in Table \ref{tab:Tracker_param}.
\begin{table}[htb]
\centering
\caption{IE-PHPMHT, PMBM, and JIPDA tuning  parameter settings}
\label{tab:Tracker_param}
\renewcommand{\arraystretch}{1.3}  
\begin{tabularx}{\columnwidth}{lll}
\hline
\hline
\textbf{Parameter} & \textbf{Symbol} & \textbf{Value} \\
\hline
Common Parameters & & \\
\hline
Sample interval & $T$ & 2.5\,s \\
Process noise intensity & $q$ & 0.01\,m$^2$s$^{-3}$ \\
Existence termination threshold & $t_d$ & $10^{-3}$ \\
\hline
IE-PHPMHT Part & & \\
\hline
Target spread & $\sigma_r, \ \sigma_\theta$ & $37.5$\,m, $1$\,$^\circ$ \\
Initial Gamma distr. shape parameter & $\alpha$ & 20 \\
Initial Gamma distr. rate parameter & $\beta$ & 1 \\
Initial Exponential distr. rate parameter & $\gamma$ & 500 \\
Probability of survival & $p_s$ & 0.95 \\
Probability of birth & $p_b$ & $10^{-4}$ \\
Existence confirmation threshold & $t_c$ & 0.5 \\
Adaptive birth detection threshold & $\tau_\text{L}$ & 0.12 \\
Adaptive birth PMBM proximity & $\varepsilon$ & 100\,m \\
\hline
PMBM Part & & \\
\hline
Detection threshold & $\tau_\text{H}$ & 0.9 \\
Probability of survival & $p_s$ & 0.999 \\
Detection probability & $p_d$ & 0.9 \\
Clutter intensity & $\mu^\text{c}$ & $10^{-6}\,\text{m}^{-2}$ \\
Birth weight & $w_b$ & 0.1 \\
Bernoulli pruning threshold & $t_{\text{Bp}}$ & $10^{-4}$ \\
Poisson pruning threshold & $t_{\text{Pp}}$ & $10^{-5}$ \\
Existence estimation threshold & $t_e$ & 0.5 \\
Gating threshold & $g$ & 30 \\
Max. number of hypotheses & $N_{\text{max}}$ & 200 \\
\hline
JIPDA & & \\
\hline
Detection threshold & $\tau_\text{H}, \tau_\text{L}$ & 0.9, 0.12 \\
Clutter intensity & $\mu^\text{c}$ & $10^{-6}\,\text{m}^{-2}$ \\
Detection probability & $p_d$ & 0.9 \\
Existence confirmation threshold & $t_c$ & 0.5, 0.99 \\
Gating threshold & $g$ & 30 \\
Max. number of tracks & $N_{\text{max}}$ & 200 \\
\hline
\end{tabularx}
\end{table}

The following kinematic and measurement models are used. Each target inside the FoV is represented by the state vector $x_\timeind^{m} = \left[ p_{x,\timeind}^{m} \ v_{x,\timeind}^{m} \ p_{y,\timeind}^{m} \ v_{y,\timeind}^{m} \right]^{\transp}$, where $p_{x,\timeind}^{m}$ and $p_{y,\timeind}^{m}$ denote the Cartesian position coordinates, and $v_{x,\timeind}^{m}$ and $v_{y,\timeind}^{m}$ represent the corresponding velocity components.
Assuming a target survives between time steps $\timeind - 1$ and $\timeind$, its motion evolves independently of other targets according to the state transition density, i.e. $p(x_\timeind^{m} | x_{\timeind-1}^{m}) = \mathcal{N}(x_\timeind^{m};\, F x_{\timeind-1}^{m},\, Q)$, a linear-Gaussian kinematic model. Under the nearly constant velocity (CV) model~\cite{bar-shalom_estimation_2001}, the transition matrix $F$ and the process noise covariance $Q$ are given as:
\begin{align}
F &= I_2 \otimes \begin{bmatrix} 1 & T \\ 0 & 1 \end{bmatrix}, &
Q &= q\, I_2 \otimes \begin{bmatrix} \frac{1}{3}T^3 & \frac{1}{2}T^2 \\ \frac{1}{2}T^2 & T \end{bmatrix},
\end{align}
where $I_2$ is the $2\times2$ identity matrix, $\otimes$ is the Kronecker product, $q = 0.5\,\text{m}^2\text{s}^{-3}$ is the process noise intensity, and $T = 2.5\,\mathrm{s}$ is the sampling interval.

Radar measurements are obtained in polar coordinates \((r, \theta)\), corresponding to range and azimuth. The measurement model is therefore defined as
$
p^{m}(y | x_k^{m}) = \mathcal{N}(y;\, h(x_k^{m}), R)
$,
where the measurement function \(h(\cdot)\) maps Cartesian state to polar measurements:
\begin{equation}
h(x_k^{m}) =
\begin{bmatrix}
\sqrt{\left(p_{x,k}^{m})^2 + (p_{y,k}^{m}\right)^2} \ \ \\
\arctan \left(p_{y,k}^{m} / p_{x,k}^{m}\right)
\end{bmatrix},
\end{equation}
and $R = \operatorname{diag}([\sigma_r^2, \sigma_\theta^2])$ is the measurement noise covariance. Note that $y$ is the notation for the IE-PHPMHT, however, the same model is valid for the point-based method when denoting the measurements with $c$ instead. 

Since the number of hypotheses and Bernoulli components in the PMBM filter increases with every time step, we use complexity management that follows \cite{garcia-fernandez_poisson_2018}. After each update step, several global hypotheses are maintained, each containing Bernoulli components representing potentially detected targets.
To lower the number of global hypotheses, we further restrict the number of Bernoulli components created at each time step by gating measurements within the gating threshold $g$ and select the best hypotheses using Murty's algorithm. 
Additionally, not all Bernoulli components are propagated to the next time step, and only the top global hypotheses based on weights are maintained, as in \cite{garcia-fernandez_poisson_2018}. Bernoulli components with existence probability below a threshold $t_\text{Bp}$, and those not present in any retained hypothesis, are discarded. Likewise, Poisson densities with weights below a threshold $t_\text{Pp}$ are also pruned. Lastly, the overall number of hypotheses is limited to $N_\textbf{max}$. Of the estimation methods proposed in \cite{garcia-fernandez_poisson_2018}, we use the first: selecting the global hypothesis with the highest weight. From this hypothesis, all Bernoulli components with existence probability greater than a threshold $t_e$ are chosen as output targets.

\begin{figure}[h]
    \centering
    \includegraphics[width=0.9\linewidth]{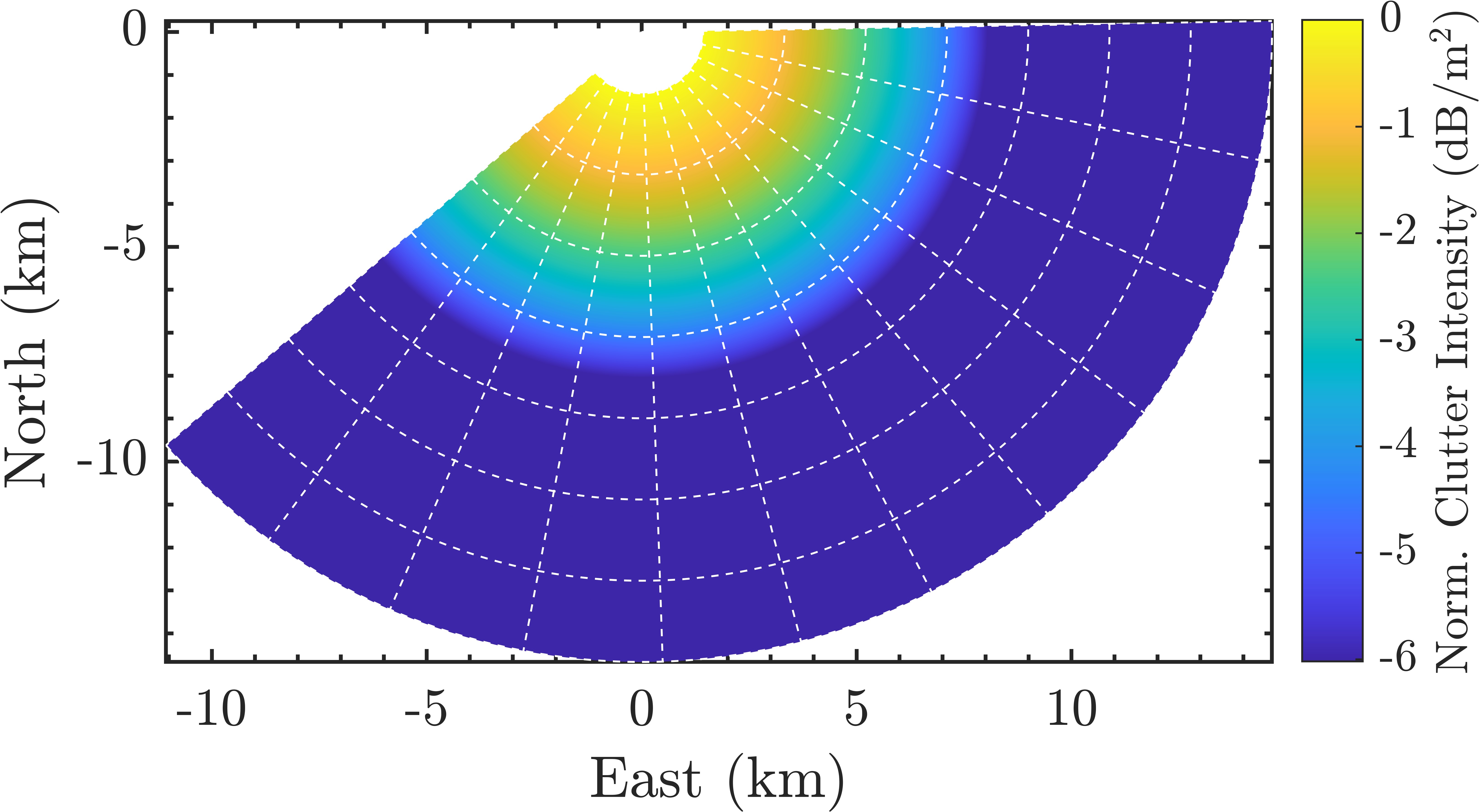}
    \caption{Visualisation of the normalised clutter intensity with the range-dependent spatial clutter distribution $p^0(y)$ used in the IE-PHPMHT.}
    \label{fig:clutter_intensity}
\end{figure}

In the conventional H-PMHT framework, the spatial clutter density $p^0(y)$ in \eqref{eq:meas_mod_intensity} is typically modelled as uniform across the surveillance region. However, in our radar application, we have noted that the characteristics of the radar and electromagnetic wave propagation result in higher clutter originating from the sea surface at shorter ranges. Therefore, we adopt a spatial clutter density that varies with range, as illustrated in Fig.~\ref{fig:clutter_intensity}.

\begin{figure*}[htb]
    \centering
    \includegraphics[width=0.329\textwidth]{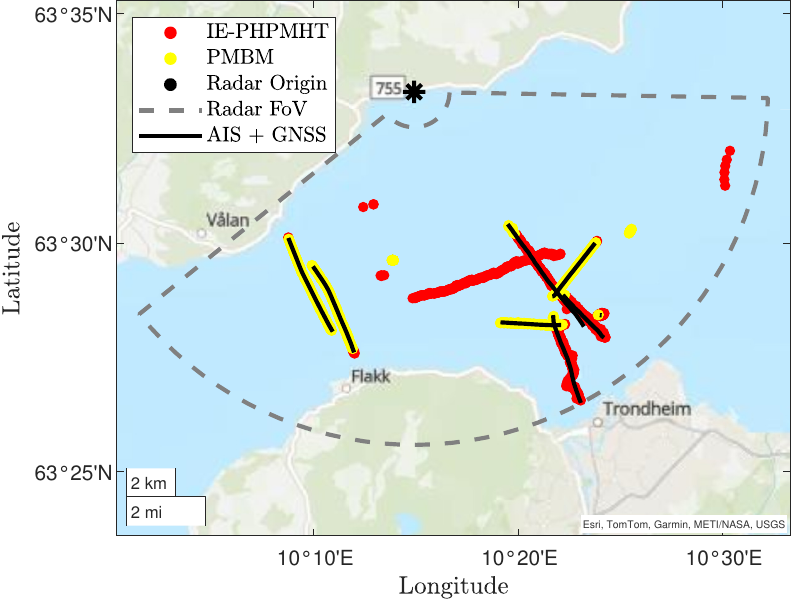}
    \includegraphics[width=0.329\textwidth]{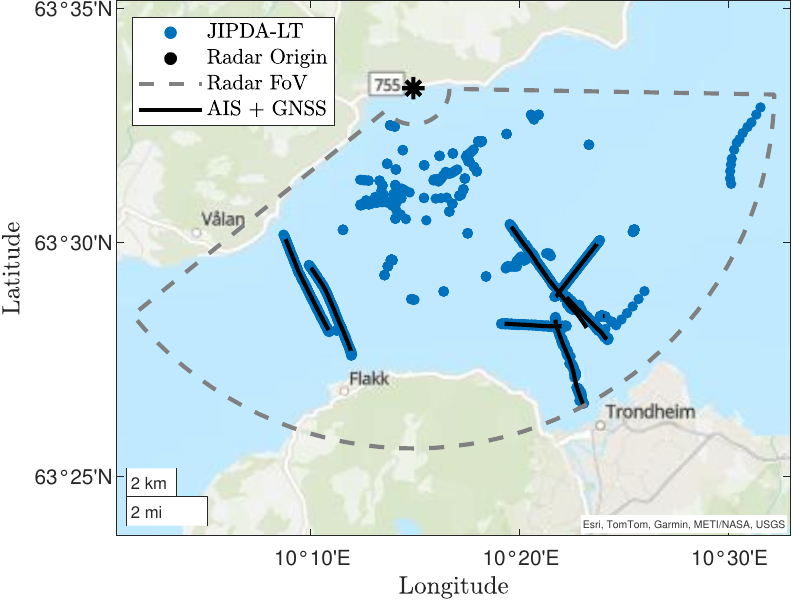}
    \includegraphics[width=0.329\textwidth]{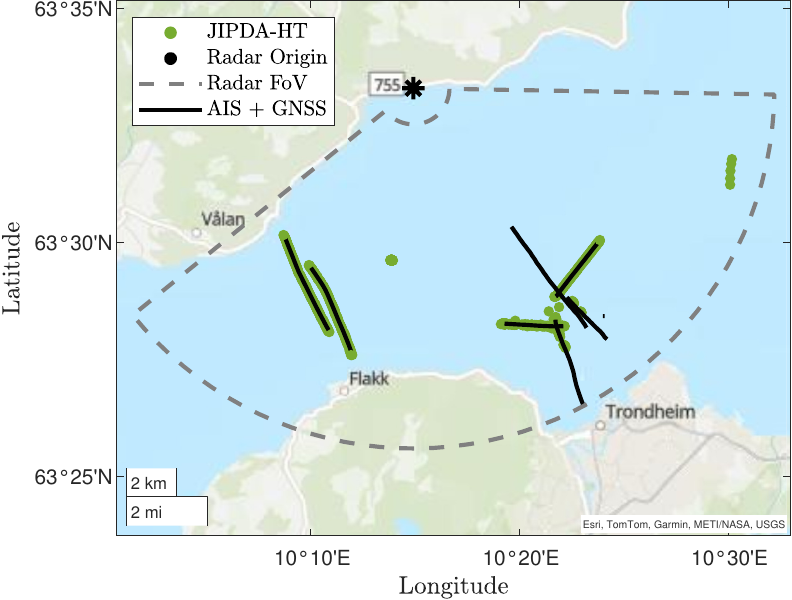}
    \caption{Tracking results comparison between the fused hybrid PMBM+IE-PHPMHT (left), low-threshold JIPDA (centre), and high-threshold JIPDA (right). Ground truth (AIS+GNSS) is shown in black. In the hybrid approach, PMBM tracks appear in yellow and IE-PHPMHT in red, closely following the ground truth with minimal false tracks. The JIPDA-LT produces many false tracks due to the low threshold, while the high threshold causes the JIPDA-HT to miss vessels.}
    \label{fig:Traj_fused_all}
\end{figure*}


The JIPDA tracker is configured with the following set of parameters. For the comparison, two different thresholding strategies are used to generate the point measurements supplied to the tracker.
First, a low threshold $\tau_\mathrm{L} = 0.12$ is used, which matches the threshold employed for the adaptive birth process in the IE-PHPMHT filter. This configuration is referred to as JIPDA-LT in the remainder.
Second, a high threshold $\tau_\mathrm{H} = 0.9$ is applied, consistent with the threshold used for point measurement extraction in the PMBM filter. This configuration is referred to as JIPDA-HT.

\subsection{Experimental Results}
\label{sec:Eval_results}



The radar dataset was processed using the hybrid PMBM+IE-PHPMHT method, and the resulting tracks were compared against those obtained from the two JIPDA variants. In particular, the evaluation of the results focuses on three key aspects: overall tracking performance, detection capabilities under low SNR conditions, and computational complexity.
\subsubsection{Overall performance}
Tracking results for all time steps are presented alongside AIS and GNSS ground truth trajectories in Fig. \ref{fig:Traj_fused_all}. In the left figure, the fused output is visually decomposed, with the PMBM component depicted in yellow and the IE-PHPMHT component in red.
It is observed that the two large car ferries, the cargo vessel, the sailing boat, and the stationary fishing boat are primarily tracked by the PMBM filter, due to their relatively high RCS. Additionally, two other stationary fishing vessels, without AIS information, are also tracked by the PMBM component.
Targets with lower RCS are generally handled by the IE-PHPMHT filter. However, when such targets approach the radar and their RCS exceeds the higher detection threshold, tracking is taken over by the PMBM filter. Another fishing vessel, also without AIS, is detected and tracked by the IE-PHPMHT in the central region of the FoV.
The track initiated by the IE-PHPMHT near the eastern edge of the FoV is likely attributable to an aircraft because of its high velocity and proximity to the airport. Furthermore, the fused results include two false tracks, persisting for two time steps, initiated by the IE-PHPMHT filter.

The central figure presents the tracking results obtained using the JIPDA-LT. It can be seen that the majority of targets are successfully tracked. However, in comparison to the PMBM+IE-PHPMHT, the continuity of tracks for targets with lower signal-to-noise ratio (SNR) is slightly worse, and it takes longer before the track on the \say{Pioner} target vessel is initiated and successfully tracked.
The primary limitation in this case, however, is the increased number of clutter tracks with short lifespans as a consequence of using a low detection threshold to achieve the observed tracking performance.

The figure on the right-hand side displays the tracking results produced by the JIPDA-HT. In this case, no erroneous clutter tracks are observed. However, the use of a higher detection threshold leads to a notable reduction in tracking performance. Specifically, the targets \say{Pioner} and \say{Viking} are not detected or tracked at any point, while \say{Elly} and \say{LOS 098} are only reliably tracked in closer ranges to the radar.

\begin{figure}[htb]
    \centering
    \includegraphics[width=\linewidth]{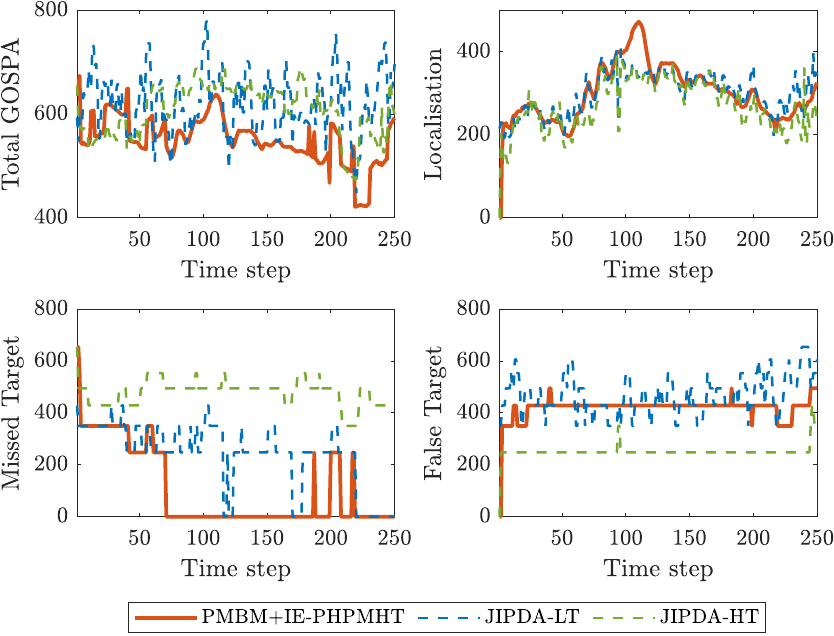}
    \caption{Comparison of the GOSPA errors and their decompositions.}
    \label{fig:GOSPA_decomp_all}
\end{figure}
For a more quantitative assessment, the performance of the tracking methods is evaluated using the \textit{generalised optimal subpattern assignment} (GOSPA) metric~\cite{rahmathullah_generalized_2017}, with parameters set to $\alpha=2$, $p=2$, and $c=350\,\text{m}$.
Fig.~\ref{fig:GOSPA_decomp_all} presents the total GOSPA error and its decomposition into localisation, missed target, and false target components for all methods. Among the compared methods, the hybrid PMBM+IE-PHPMHT approach yields the lowest total GOSPA error, followed by JIPDA-LT, while JIPDA-HT exhibits the highest overall error.
In terms of localisation error, all three methods perform comparably. This consistency is likely because of factors such as radar resolution, the digitisation process, pre-processing steps, and interpolation mismatches with the ground truth, rather than differences in tracker accuracy. An increase in localisation error is observed for the PMBM+IE-PHPMHT around time step 100, associated with a less accurate but continuous track. In contrast, the other methods fail to maintain a track at that time, thereby not contributing to the localisation component of the error.
For the missed target error, the PMBM+IE-PHPMHT shows considerably better performance than the other two methods, showing no missed targets for approximately two-thirds of the experiment duration. This highlights the advantage of the TBD approach, particularly as the JIPDA-LT receives the same point detections used in the adaptive birth process of the IE-PHPMHT. The JIPDA-HT has the highest missed target error, due to its high detection threshold, leading to missed detections of low SNR targets.
At this cost, the JIPDA-HT achieves the lowest false target error, followed by the PMBM+IE-PHPMHT. The JIPDA-LT produces the highest false target error, with noticeable fluctuations suggesting multiple short false tracks or estimates falling outside the cut-off region. It is important to note that the GOSPA evaluation is based on AIS and GNSS data, which may exclude some targets, thereby contributing to elevated false target errors for methods detecting such objects.
The average GOSPA scores across the duration of the experiment are summarised in Table \ref{tab:GOSPA}.
\begin{table}[ht]
\centering
\caption{Comparison of GOSPA averaged over all time steps}
\renewcommand{\arraystretch}{1.5}
\begin{tabular}{l||c|c|c|c}
\hline
\hline
 \textbf{Method / GOSPA}  & \textbf{Total} & \textbf{Localisation} & \textbf{Missed} & \textbf{False} \\
\hline
PMBM+IE-PHPMHT & \underline{550.86} & 299.03 & 102.30 & 418.18 \\
\hline
JIPDA-LT & 621.11 & 299.63 & 240.24 & 466.98 \\
\hline
JIPDA-HT & 606.51 & 270.33 & 477.18 & 250.41 \\
\hline
\end{tabular}
\label{tab:GOSPA}
\end{table}

\begin{figure}[htb]
    \centering
    \includegraphics[width=0.9\linewidth]{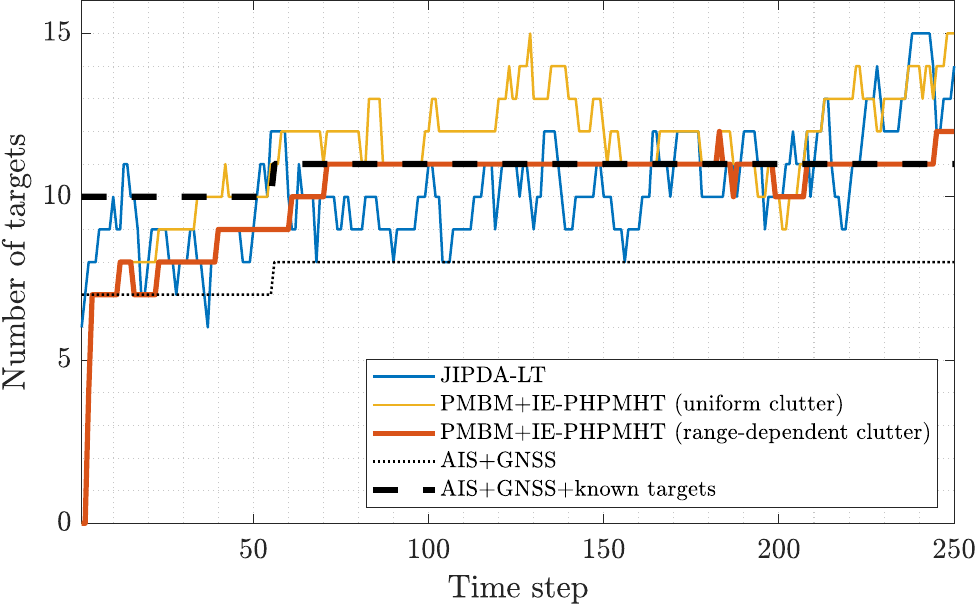}
    \caption{Estimated number of targets over time for the different tracking methods. Additional visually confirmed targets, not present in the AIS or GNSS ground truth, are included. The hybrid PMBM+IE-PHPMHT with the range-dependent spatial clutter distribution shows stable estimates with minor deviations, while the JIPDA-LT exhibits greater offset and variation in the estimated target count over time.}
    \label{fig:cardinality_comp}
\end{figure}

Fig. \ref{fig:cardinality_comp} shows the estimated number of targets over time. In this plot, additional known targets present within the FoV, but not included in the AIS and GNSS ground truth logs, have been taken into account. The presence of these additional targets was confirmed through visual observation during the experiment and is considered only in this cardinality analysis, as their precise locations are not required. The hybrid PMBM+IE-PHPMHT method correctly estimates the number of targets from time step 70 onwards, except for four short instances of false or missed targets, whereas the latter is attributed to the presumed aircraft entering the scene.
This result also highlights the effectiveness of the range-dependent clutter distribution (see Fig. \ref{fig:clutter_intensity}) utilised in the IE-PHPMHT, which yields a more stable and accurate estimation of target cardinality. In contrast, the JIPDA-LT tends to underestimate the number of targets and exhibits notable fluctuations throughout the experiment. These variations are primarily due to false tracks and track fragmentation, which can also be seen in Fig. \ref{fig:Traj_fused_all}.

\subsubsection{Low SNR detection capabilities}
\begin{figure}[htb]
    \centering
    \includegraphics[width=0.9\linewidth]{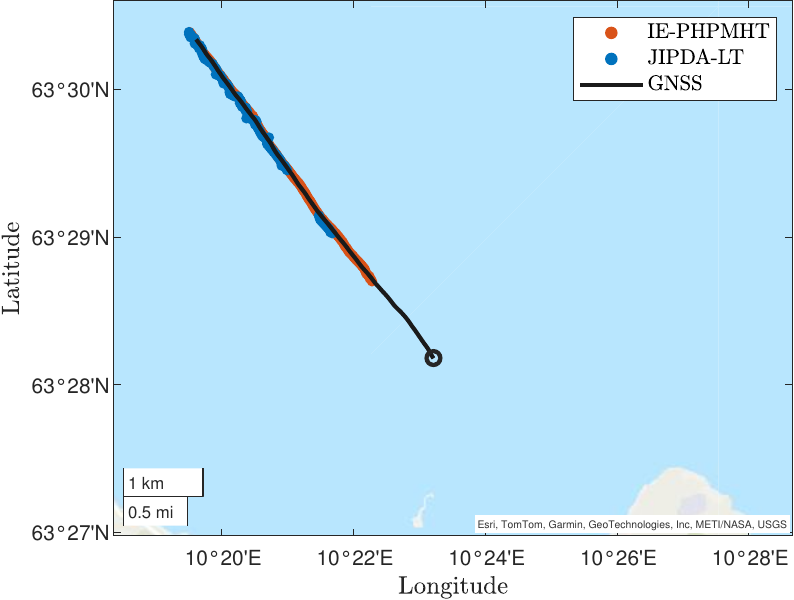}
    \caption{Comparison of IE-PHPMHT and JIPDA-LT tracking performance for the low-observable target Pioner. The circle marks the starting position. Other ground truth tracks and estimates are omitted for clarity.}
    \label{fig:Traj_TBD_comp}
\end{figure}

Next, we analyse the detection capabilities of the IE-PHPMHT TBD approach in comparison to the detection-based JIPDA tracker. Specifically, we examine how early each method is able to establish a track on the low-observable target \say{Pioner}, which starts at time step 1 and moves towards the radar from an initial distance of approximately 9\,km. The tracking results of IE-PHPMHT and JIPDA-LT are illustrated in Fig.~\ref{fig:Traj_TBD_comp}, alongside the GNSS-based ground truth.
The IE-PHPMHT algorithm successfully detects and establishes a track at time step 99, while the JIPDA-LT tracker establishes a consistent track later, at time step 151. It should be noted that JIPDA-LT briefly produces a short-lived track segment around time step 110, possibly due to fluctuations, but fails to maintain continuity.
In this scenario, the approximate SNR at which reliable tracking becomes possible is about 8\,dB for IE-PHPMHT and 10\,dB for JIPDA-LT. The resulting 2\,dB gain in detection sensitivity enables the TBD method to establish the track approximately two minutes earlier, corresponding to a range advantage of about 2\,km.

\subsubsection{Computational complexity}
Finally, the computational cost of the proposed method is assessed through an analysis and comparison of the runtimes. The average runtimes per time step are given in Table \ref{tab:runtime}. The radar frames processed have a resolution of 511$\times$1711 cells. The PMBM filter and the JIPDA-HT, both operating with the higher detection threshold, have the lowest computational demands. Reducing the threshold results in an increased number of point measurements, thereby increasing the runtime for the JIPDA-LT variant.
As expected, the IE-PHPMHT, operating directly on the full radar image without a pre-detection stage, has the longest runtime. Nevertheless, when combined with the PMBM tracker, the runtime of the IE-PHPMHT component is reduced, due to a decreased number of Bernoulli components to process. The additional computational burden introduced by the PMBM component remains relatively minor, such that the combined runtime is lower.
Thus, the hybrid TBD system achieves an average runtime of 1.06\,s per time step, and given a radar update interval of 2.5\,s, the system satisfies real-time processing requirements.
\begin{table}[htb]
\centering
\caption{Comparison of the average runtime per time step}
\renewcommand{\arraystretch}{1.3}
\begin{tabular}{l|c}
\hline
\hline
\textbf{Method} & \textbf{Time (s)} \\
\hline
PMBM & 0.14 \\
IE-PHPMHT & 1.21 \\
PMBM+IE-PHPMHT & 1.06 \\
JIPDA-LT & 0.28 \\
JIPDA-HT & 0.18 \\
\hline
\end{tabular}
\label{tab:runtime}
\end{table}

%% file: 05_Conclusion.tex
\section{Conclusion}
\label{sec:conclusion}
In this paper, we have developed a hybrid multi-target TBD tracking system for application to coastal maritime surveillance to address the need to robustly track multiple maritime targets, including small, low-RCS vessels, in large, cluttered and dynamic environments with limited computational burden.
The hybrid approach combines a detection-based PMBM filter with a TBD IE-PHPMHT, leveraging unthresholded raw radar data, to balance the trade-off between detection performance and computational real-time demands.
A full-scale implementation of the complete signal processing chain has been presented and evaluated using real-world X-band radar data collected during experiments conducted in the Trondheim Fjord, Norway. We have demonstrated that large-scale TBD can be implemented to meet real-time requirements.
In the system evaluation, experimental results showed that the proposed hybrid tracker successfully maintained continuous tracking of both high- and low-RCS targets throughout the experiment. Comparative evaluations with high- and low-threshold JIPDA trackers demonstrated superior performance in terms of target detection while maintaining a low number of false tracks, as quantified by the GOSPA metric.

In future work, other hybrid topologies such as the one in \cite{kropfreiter_multiobject_2024} could be explored. It is also relevant to analyse exploiting the benefit of sharing information on a deeper level (e.g. Bernoulli components) than the track level between the PMBM filter and the TBD tracking algorithm. Another line of future research is the extension to multi-sensor applications to have a wider coverage and mitigate the effect of shadowing.

%% file: 00_Main.bbl

%% file: 00_Main.bbl
\begin{thebibliography}{10}
\providecommand{\url}[1]{#1}
\csname url@samestyle\endcsname
\providecommand{\newblock}{\relax}
\providecommand{\bibinfo}[2]{#2}
\providecommand{\BIBentrySTDinterwordspacing}{\spaceskip=0pt\relax}
\providecommand{\BIBentryALTinterwordstretchfactor}{4}
\providecommand{\BIBentryALTinterwordspacing}{\spaceskip=\fontdimen2\font plus
\BIBentryALTinterwordstretchfactor\fontdimen3\font minus \fontdimen4\font\relax}
\providecommand{\BIBforeignlanguage}[2]{{%
\expandafter\ifx\csname l@#1\endcsname\relax
\typeout{** WARNING: IEEEtran.bst: No hyphenation pattern has been}%
\typeout{** loaded for the language `#1'. Using the pattern for}%
\typeout{** the default language instead.}%
\else
\language=\csname l@#1\endcsname
\fi
#2}}
\providecommand{\BIBdecl}{\relax}
\BIBdecl

\bibitem{maresca_maritime_2014}
S.~Maresca, P.~Braca, J.~Horstmann, and R.~Grasso, ``Maritime {Surveillance} {Using} {Multiple} {High}-{Frequency} {Surface}-{Wave} {Radars},'' \emph{IEEE Trans. Geosci. Remote Sens.}, vol.~52, no.~8, pp. 5056--5071, Aug. 2014.

\bibitem{vivone_joint_2016}
G.~Vivone and P.~Braca, ``Joint {Probabilistic} {Data} {Association} {Tracker} for {Extended} {Target} {Tracking} {Applied} to {X}-{Band} {Marine} {Radar} {Data},'' \emph{IEEE J. Ocean. Eng.}, vol.~41, no.~4, pp. 1007--1019, Oct. 2016.

\bibitem{forti_next-gen_2022}
N.~Forti, E.~d'Afflisio, P.~Braca, L.~M. Millefiori, S.~Carniel, and P.~Willett, ``Next-{Gen} {Intelligent} {Situational} {Awareness} {Systems} for {Maritime} {Surveillance} and {Autonomous} {Navigation},'' \emph{Proceedings of the IEEE}, vol. 110, no.~10, pp. 1532--1537, Oct. 2022.

\bibitem{brekke_autosea_2019}
E.~F. Brekke, E.~F. Wilthil, B.-O.~H. Eriksen, D.~K.~M. Kufoalor, O.~K. Helgesen, I.~B. Hagen, M.~Breivik, and T.~A. Johansen, ``\BIBforeignlanguage{en}{The {Autosea} project: {Developing} closed-loop target tracking and collision avoidance systems},'' \emph{\BIBforeignlanguage{en}{J. Phys.: Conf. Ser.}}, vol. 1357, no. 012020, Oct. 2019.

\bibitem{bar-shalom_estimation_2001}
Y.~Bar-Shalom, X.-R. Li, and T.~Kirubarajan, \emph{Estimation with {Applications} to {Tracking} and {Navigation}}.\hskip 1em plus 0.5em minus 0.4em\relax New York (NY): Wiley, 2001.

\bibitem{richards_fundamentals_2014}
M.~A. Richards, \emph{\BIBforeignlanguage{en}{Fundamentals of {Radar} {Signal} {Processing}}}.\hskip 1em plus 0.5em minus 0.4em\relax New York (NY): McGraw-Hill Education, 2014.

\bibitem{reid_algorithm_1979}
D.~Reid, ``An algorithm for tracking multiple targets,'' \emph{IEEE Trans. Autom. Control.}, vol.~24, no.~6, pp. 843--854, Dec. 1979.

\bibitem{mellema_improved_2020}
G.~R. Mellema, ``Improved {Active} {Sonar} {Tracking} in {Clutter} {Using} {Integrated} {Feature} {Data},'' \emph{IEEE J. Ocean. Eng.}, vol.~45, no.~1, pp. 304--318, Jan. 2020.

\bibitem{fortmann_multi-target_1980}
T.~E. Fortmann, Y.~Bar-Shalom, and M.~Scheffe, ``Multi-target tracking using joint probabilistic data association,'' in \emph{Proc. {IEEE} {Conf}. {Decis}. {Control}.}, Dec. 1980.

\bibitem{fortmann_sonar_1983}
T.~Fortmann, Y.~Bar-Shalom, and M.~Scheffe, ``Sonar tracking of multiple targets using joint probabilistic data association,'' \emph{IEEE J. Ocean. Eng.}, vol.~8, no.~3, pp. 173--184, Jul. 1983.

\bibitem{mahler_advances_2014}
R.~P.~S. Mahler, Ed., \emph{\BIBforeignlanguage{eng}{Advances in {Statistical} {Multisource}-{Multitarget} {Information} {Fusion}}}.\hskip 1em plus 0.5em minus 0.4em\relax Norwood (MA): Artech House, 2014.

\bibitem{williams_marginal_2015}
J.~L. Williams, ``Marginal multi-bernoulli filters: {RFS} derivation of {MHT}, {JIPDA}, and association-based {MeMBer},'' \emph{IEEE Trans. Aerosp. Electron. Syst.}, vol.~51, no.~3, pp. 1664--1687, Jul. 2015.

\bibitem{garcia-fernandez_poisson_2018}
A.~F. García-Fernández, J.~L. Williams, K.~Granström, and L.~Svensson, ``Poisson {Multi}-{Bernoulli} {Mixture} {Filter}: {Direct} {Derivation} and {Implementation},'' \emph{IEEE Trans. Aerosp. Electron. Syst.}, vol.~54, no.~4, pp. 1883--1901, Aug. 2018.

\bibitem{mallick_integrated_2013}
M.~Mallick, V.~Krishnamurthy, and B.-N. Vo, \emph{\BIBforeignlanguage{en}{Integrated {Tracking}, {Classification}, and {Sensor} {Management}: {Theory} and {Applications}}}.\hskip 1em plus 0.5em minus 0.4em\relax Hoboken (NJ): John Wiley \& Sons Inc, 2013.

\bibitem{davey_comparison_2008}
S.~Davey, M.~Rutten, and B.~Cheung, ``A comparison of detection performance for several {Track}-{Before}-{Detect} algorithms,'' in \emph{Proc. 11th {Int}. {Conf}. {Inf}. {Fusion} ({FUSION})}, Jun. 2008.

\bibitem{davies_information_2024}
E.~S. Davies and A.~F. García-Fernández, ``\BIBforeignlanguage{en}{Information {Exchange} {Track}-{Before}-{Detect} {Multi}-{Bernoulli} {Filter} for {Superpositional} {Sensors}},'' \emph{\BIBforeignlanguage{en}{IEEE Trans. Signal Process.}}, vol.~72, pp. 607--621, 2024.

\bibitem{garcia-fernandez_track-before-detect_2016}
A.~F. García-Fernández, ``Track-before-detect labeled multi-{Bernoulli} particle filter with label switching,'' \emph{IEEE Trans. Aerosp. Electron. Syst.}, vol.~52, no.~5, pp. 2123--2138, Oct. 2016.

\bibitem{ristic_bernoulli_2020}
B.~Ristic, L.~Rosenberg, D.~Y. Kim, X.~Wang, and J.~Williams, ``\BIBforeignlanguage{en}{Bernoulli track-before-detect filter for maritime radar},'' \emph{\BIBforeignlanguage{en}{IET Radar Sonar Navig.}}, vol.~14, no.~3, pp. 356--363, 2020.

\bibitem{vo_joint_2010}
B.-N. Vo, B.-T. Vo, N.-T. Pham, and D.~Suter, ``\BIBforeignlanguage{en}{Joint {Detection} and {Estimation} of {Multiple} {Objects} {From} {Image} {Observations}},'' \emph{\BIBforeignlanguage{en}{IEEE Trans. Signal Process.}}, vol.~58, no.~10, pp. 5129--5141, Oct. 2010.

\bibitem{barniv_dynamic_1985}
Y.~Barniv, ``Dynamic {Programming} {Solution} for {Detecting} {Dim} {Moving} {Targets},'' \emph{IEEE Trans. Aerosp. Electron. Syst.}, vol. AES-21, no.~1, pp. 144--156, Jan. 1985.

\bibitem{yi_efficient_2013}
W.~Yi, M.~R. Morelande, L.~Kong, and J.~Yang, ``An {Efficient} {Multi}-{Frame} {Track}-{Before}-{Detect} {Algorithm} for {Multi}-{Target} {Tracking},'' \emph{IEEE J. Sel. Top. Signal Process.}, vol.~7, no.~3, pp. 421--434, Jun. 2013.

\bibitem{salmond_particle_2001}
D.~Salmond and H.~Birch, ``\BIBforeignlanguage{en}{A particle filter for track-before-detect},'' in \emph{\BIBforeignlanguage{en}{Proc. {American} {Cont}. {Conf}.}}, vol.~6, 2001, pp. 3755--3760.

\bibitem{rutten_recursive_2005}
M.~Rutten, N.~Gordon, and S.~Maskell, ``\BIBforeignlanguage{en}{Recursive track-before-detect with target amplitude fluctuations},'' \emph{\BIBforeignlanguage{en}{IEE Proc., Radar Sonar Navig.}}, vol. 152, no.~5, p. 345, 2005.

\bibitem{morelande_bayesian_2007}
M.~R. Morelande, C.~M. Kreucher, and K.~Kastella, ``A {Bayesian} {Approach} to {Multiple} {Target} {Detection} and {Tracking},'' \emph{IEEE Trans. Signal Process.}, vol.~55, no.~5, pp. 1589--1604, May 2007.

\bibitem{northardt_track-before-detect_2019}
T.~Northardt and S.~C. Nardone, ``Track-{Before}-{Detect} {Bearings}-{Only} {Localization} {Performance} in {Complex} {Passive} {Sonar} {Scenarios}: {A} {Case} {Study},'' \emph{IEEE J. Ocean. Eng.}, vol.~44, no.~2, pp. 482--491, Apr. 2019.

\bibitem{streit_tracking_2000}
R.~L. Streit, ``\BIBforeignlanguage{en}{Tracking on {Intensity}-{Modulated} {Data} {Streams}},'' NUWC, Tech. Rep. 11221, May 2000.

\bibitem{davey_track-before-detect_2018}
S.~J. Davey and H.~X. Gaetjens, \emph{\BIBforeignlanguage{en}{Track-{Before}-{Detect} {Using} {Expectation} {Maximisation}}}.\hskip 1em plus 0.5em minus 0.4em\relax Singapore: Springer Singapore, 2018.

\bibitem{herrmann_histogram-probabilistic_2025}
\BIBentryALTinterwordspacing
L.~Herrmann, A.~F. García-Fernández, and E.~F. Brekke, ``\BIBforeignlanguage{en}{Histogram-{Probabilistic} {Multi}-{Hypothesis} {Tracking} with {Integrated} {Target} {Existence}},'' \emph{\BIBforeignlanguage{en}{IEEE Trans. Aerosp. Electron. Syst.}}, 2025, currently under review. [Online]. Available: \url{http://arxiv.org/abs/2504.20526}
\BIBentrySTDinterwordspacing

\bibitem{mcdonald_track-before-detect_2008}
M.~McDonald and B.~Balaji, ``\BIBforeignlanguage{en}{Track-before-{Detect} {Using} {Swerling} 0, 1, and 3 {Target} {Models} for {Small} {Manoeuvring} {Maritime} {Targets}},'' \emph{\BIBforeignlanguage{en}{EURASIP J. Adv. Signal Process.}}, vol. 2008, no.~1, pp. 1--9, Dec. 2008.

\bibitem{mcdonald_impact_2011}
------, ``Impact of {Measurement} {Model} {Mismatch} on {Nonlinear} {Track}-{Before}-{Detect} {Performance} for {Maritime} {RADAR} {Surveillance},'' \emph{IEEE J. Ocean. Eng.}, vol.~36, no.~4, pp. 602--614, Oct. 2011.

\bibitem{davey_detecting_2011}
S.~J. Davey, ``Detecting a {Small} {Boat} using {Histogram} {PMHT}.'' \emph{J. Adv. Inf. Fusion}, vol.~6, no.~2, pp. 167--186, 2011.

\bibitem{guo_sa-hpmht_2023}
Y.~Guo, K.~Teng, and L.~Shi, ``\BIBforeignlanguage{en}{{SA}-{HPMHT} for {Maritime} {Dim} {Targets} {Tracking} {With} {Sensor} {Location} {Uncertainty}},'' \emph{\BIBforeignlanguage{en}{IEEE Sens. J.}}, vol.~23, no.~5, pp. 5134--5145, Mar. 2023.

\bibitem{zhou_multiple-kernelized-correlation-filter-based_2022}
Y.~Zhou, H.~Su, S.~Tian, X.~Liu, and J.~Suo, ``Multiple-{Kernelized}-{Correlation}-{Filter}-{Based} {Track}-{Before}-{Detect} {Algorithm} for {Tracking} {Weak} and {Extended} {Target} in {Marine} {Radar} {Systems},'' \emph{IEEE Trans. Aerosp. Electron. Syst.}, vol.~58, no.~4, pp. 3411--3426, Aug. 2022.

\bibitem{ristic_track-before-detect_2024}
B.~Ristic, D.~Y. Kim, and L.~Rosenberg, ``Track-{Before}-{Detect} for {Airborne} {Maritime} {Radar}: {Application} to {Real} {Data},'' in \emph{Proc. 27th {Int}. {Conf}. {Inf}. {Fusion} ({FUSION})}, Jul. 2024.

\bibitem{herrmann_track_2025}
L.~Herrmann, A.~F. García-Fernández, E.~F. Brekke, and E.~Eide, ``Track {Initiation} and {Adaptive} {Target} {Birth} in {Existence}-{Based} {Poisson} {Histogram}-{PMHT},'' in \emph{to appear in {Proc}. {IEEE} {Radar} {Conf}.}, Oct. 2025.

\bibitem{ward_sea_2013}
K.~Ward, R.~Tough, and S.~Watts, \emph{\BIBforeignlanguage{en}{Sea {Clutter}: {Scattering}, the {K} {Distribution} and {Radar} {Performance}}}.\hskip 1em plus 0.5em minus 0.4em\relax IET Digital Library, 2013.

\bibitem{herrmann_coherent_2024}
L.~Herrmann, E.~F. Brekke, and E.~Eide, ``Coherent {Integration} of {Optical} {Flow} for {Track}-{Before}-{Detect} {Radar} {Detection},'' in \emph{Proc. 27th {Int}. {Conf}. {Inf}. {Fusion} ({FUSION})}, Jul. 2024.

\bibitem{herrmann_target_2025}
L.~Herrmann, G.~Tabella, E.~F. Brekke, and E.~Eide, ``Target {Detection} in {Maritime} {Radar} {Tracking} {Based} on {Spatial} {Image} {Gradients},'' \emph{IEEE Sens. J.}, vol.~25, no.~13, pp. 24\,886--24\,897, Jul. 2025.

\bibitem{ester_density-based_1996}
M.~Ester, H.~Kriegel, J.~Sander, and X.~Xu, ``A {Density}-{Based} {Algorithm} for {Discovering} {Clusters} in {Large} {Spatial} {Databases} with {Noise},'' in \emph{Proceedings of the {Second} {International} {Conference} on {Knowledge} {Discovery} and {Data} {Mining}}, 1996.

\bibitem{bishop_pattern_2006}
C.~M. Bishop, \emph{\BIBforeignlanguage{en}{Pattern {Recognition} and {Machine} {Learning}}}.\hskip 1em plus 0.5em minus 0.4em\relax New York (NY): Springer, 2006.

\bibitem{brekke_relationship_2018}
E.~Brekke and M.~Chitre, ``\BIBforeignlanguage{en}{Relationship {Between} {Finite} {Set} {Statistics} and the {Multiple} {Hypothesis} {Tracker}},'' \emph{\BIBforeignlanguage{en}{IEEE Trans. Aerosp. Electron. Syst.}}, vol.~54, no.~4, pp. 1902--1917, Aug. 2018.

\bibitem{granstrom_poisson_2020}
K.~Granström, M.~Fatemi, and L.~Svensson, ``Poisson {Multi}-{Bernoulli} {Mixture} {Conjugate} {Prior} for {Multiple} {Extended} {Target} {Filtering},'' \emph{IEEE Trans. Aerosp. Electron. Syst.}, vol.~56, no.~1, pp. 208--225, Feb. 2020.

\bibitem{baerveldt_multiple_2024}
M.~Baerveldt, M.~E. López, and E.~F. Brekke, ``\BIBforeignlanguage{en}{A {Multiple} {Extended} {Object} {Tracker} with the {Gaussian} {Process} {Model} {Utilizing} {Negative} {Information}},'' \emph{\BIBforeignlanguage{en}{J. Adv. Inf. Fusion}}, vol.~19, no.~1, pp. 88--108, Dec. 2024.

\bibitem{granstrom_poisson_2025}
K.~Granström, L.~Svensson, Y.~Xia, J.~Williams, and A.~F. García-Fernández, ``Poisson {Multi}-{Bernoulli} {Mixtures} for {Sets} of {Trajectories},'' \emph{IEEE Trans. Aerosp. Electron. Syst.}, vol.~61, no.~2, pp. 5178--5194, Apr. 2025.

\bibitem{musicki_joint_2004}
D.~Musicki and R.~Evans, ``\BIBforeignlanguage{en}{Joint integrated probabilistic data association: {JIPDA}},'' \emph{\BIBforeignlanguage{en}{IEEE Trans. Aerosp. Electron. Syst.}}, vol.~40, no.~3, pp. 1093--1099, Jul. 2004.

\bibitem{rahmathullah_generalized_2017}
A.~S. Rahmathullah, A.~F. García-Fernández, and L.~Svensson, ``\BIBforeignlanguage{en}{Generalized optimal sub-pattern assignment metric},'' in \emph{\BIBforeignlanguage{en}{Proc. 20th {Int}. {Conf}. {Inf}. {Fusion} ({FUSION})}}, Jul. 2017.

\bibitem{kropfreiter_multiobject_2024}
T.~Kropfreiter, J.~L. Williams, and F.~Meyer, ``Multiobject {Tracking} for {Thresholded} {Cell} {Measurements},'' in \emph{Proc. 27th {Int}. {Conf}. {Inf}. {Fusion} ({FUSION})}, Jul. 2024.

\end{thebibliography}
